\documentclass[11pt,a4paper]{article}
\pdfoutput=1
\usepackage{jcappub}
\usepackage{blindtext}
\usepackage{hyperref}
\usepackage{amsmath,amssymb}
\usepackage{microtype}
\usepackage{graphicx}
\usepackage{bm}
\usepackage{latexsym}
\usepackage{epsfig}
\usepackage{color}
\usepackage[dvipsnames]{xcolor}
\usepackage{subfigure}
\usepackage{graphicx}
\usepackage{amssymb}
\usepackage{amsmath}
\usepackage{bm}
\usepackage{latexsym}
\usepackage{epsfig}
\usepackage{amsmath}
\usepackage[normalem]{ulem}
\usepackage{textcomp}
\usepackage{color}
\usepackage[utf8]{inputenc}
\usepackage{array}
\usepackage{comment}
\usepackage{mathalpha}
\usepackage{tabularray}
\makeatletter
\gdef\@fpheader{}
\makeatother
\newcommand{\bb}[1]{{\color{red}[BB: #1]}}

\def\beq{\begin{equation}}
\def\eeq{\end{equation}}
\def\bea{\begin{eqnarray}}
\def\eea{\end{eqnarray}}
\def\be{\begin{equation}}
\def\ee{\end{equation}}

\def\bse{\begin{subequations}}
\def\ese{\end{subequations}}

\def\ee{\eta_{\rm e}}

\def\Mpl{M_{P}}
\def\f{\frac}
\def\l{\left}
\def\r{\right}

\def\Teq{T_{\text {eq}}}
\def\Tbh{T_{\text {BH}}}
\def\Tev{T_{\text {ev}}}
\def\tin{t_{\rm in}}

\def\Min{M_{\text {in}}}

\def\mbh{M_{\text {BH}}}

\def\Trh{T_{\text {RH}}}
\def\gammabh{\Gamma_{\rm{BH}}}
\def\mdm{M_{\text {DM}}}

\def\hin{H_{\rm in}}

\def\ain{a_{\rm in}}

\def\Hev{H_\text{ev}}

\def\aev{a_\text{ev}}

\def\Tin{T_\text{in}}
\def\mdm{m_\text{DM}}
\def\gss{g_{\star s}}
\def\Teq{T_\text{eq}^{\rm BH}}
\def\Neff{N_{\rm eff}}
\def\DNeff{\Delta N_{\rm eff}}

\begin{document}
\title{Gravitational wave signatures of cogenesis from a burdened PBH}
\author[a]{Basabendu Barman,}
\emailAdd{basabendu.b@srmap.edu.in}
\author[b]{Md Riajul Haque}
\emailAdd{riaj.0009@gmail.com}
\author[c]{and Óscar Zapata}
\emailAdd{oalberto.zapata@udea.edu.co}
\affiliation[a]{\,\,Department of Physics, School of Engineering and Sciences,
SRM University-AP, Amaravati 522240, India}
\affiliation[b]{\,\,Centre for Strings, Gravitation and Cosmology,
Department of Physics, Indian Institute of Technology Madras, 
Chennai~600036, India}
\affiliation[c]{\,\,Instituto de Física, Universidad de Antioquia\\Calle 70 \# 52-21, Apartado Aéreo 1226, Medellín, Colombia}
\abstract
{
We explore the possibility of explaining the observed dark matter (DM) relic abundance, along with matter-antimatter asymmetry, entirely from the evaporation of primordial black holes (PBH) beyond the semi-classical approximation. We find that, depending on the timing of modification to the semi-classical approximation and the efficiency of the backreaction, it is possible to produce the correct DM abundance for PBHs with masses $\gtrsim\mathcal{O}(10^3)$ g, whereas producing the right amount of baryon asymmetry requires light PBHs with masses $\lesssim\mathcal{O}(10^3)$ g, satisfying bounds on the PBH mass from the Cosmic Microwave Background and Big Bang Nucleosynthesis. However, in a simplistic scenario, achieving both {\it simultaneously} is not feasible, typically because of the stringent Lyman-$\alpha$ constraint on warm dark matter mass. In addition to DM and baryon asymmetry, we also investigate the impact of memory burden on dark radiation, evaporated from PBH, constrained by the effective number of relativistic degrees of freedom $\DNeff$. Furthermore, we demonstrate how induced gravitational waves from PBH density fluctuations can provide a window to test the memory-burden effects, thereby placing constraints on either the DM mass scale or the scale of leptogenesis.
}
\maketitle
\section{Introduction}
\label{sec:intro}
Originally introduced by Stephen Hawking, primordial black holes (PBHs) possess intriguing cosmic characteristics~\cite{Hawking:1974rv, Hawking:1975vcx}. PBHs with masses $M_{\rm BH} \gtrsim 10^{15}$ g persist as stable entities in the contemporary Universe and are viable candidates for dark matter (DM) (for a comprehensive overview, see, for example, Ref.~\cite{Carr:2020xqk}). Conversely, primordial black holes must be considerably lighter to investigate particle production from its evaporation. Specifically, their formation mass should fall within a range permitting evaporation prior to Big Bang Nucleosynthesis (BBN), corresponding to $M_{\rm BH} \lesssim 10^{9}$ g. Deviation from this criterion could introduce additional parameters, potentially disrupting the accurate prediction of BBN, crucially inferred from the measurement of $\DNeff$~\cite{Planck:2018vyg}. Within this mass spectrum, PBHs can decay, thereby playing a pivotal role in generating not only Standard Model (SM) particles but also exotic new physics states, {\it viz.,} DM. Various studies have explored DM production~\cite{Morrison:2018xla, Gondolo:2020uqv, Bernal:2020bjf, Green:1999yh, Khlopov:2004tn, Dai:2009hx, Allahverdi:2017sks, Lennon:2017tqq, Hooper:2019gtx, Chaudhuri:2020wjo, Masina:2020xhk, Baldes:2020nuv, Bernal:2020ili, Bernal:2020kse, Lacki:2010zf, Boucenna:2017ghj, Adamek:2019gns, Carr:2020mqm, Masina:2021zpu, Bernal:2021bbv, Bernal:2021yyb, Samanta:2021mdm, Sandick:2021gew, Cheek:2021cfe, Cheek:2021odj, Barman:2021ost, Borah:2022iym,Chen:2023lnj,Chen:2023tzd,Kim:2023ixo,Gehrman:2023qjn,Chaudhuri:2023aiv}, 
baryon asymmetry~\cite{Baumann:2007yr, Hook:2014mla, Fujita:2014hha, Hamada:2016jnq, Morrison:2018xla, Hooper:2020otu, Perez-Gonzalez:2020vnz, Datta:2020bht, JyotiDas:2021shi, Smyth:2021lkn, Barman:2021ost, Bernal:2022pue, Ambrosone:2021lsx,Calabrese:2023key,Calabrese:2023bxz,Gehrman:2022imk,Gehrman:2023esa,Schmitz:2023pfy} or cogenesis~\cite{Fujita:2014hha, Morrison:2018xla, Hooper:2019gtx, Lunardini:2019zob, Masina:2020xhk, Hooper:2020otu, Datta:2020bht, JyotiDas:2021shi, Schiavone:2021imu, Bernal:2021yyb, Bernal:2021bbv, Bernal:2022swt,Barman:2022pdo,Borah:2024qyo,Chianese:2024nyw}, both in the standard radiation dominated background, as well as in an inflaton dominated background during reheating~\cite{Haque:2024cdh,Barman:2024slw}. It is crucial to mention here that the semi-classical phenomenon of PBH evaporation heavily relies on the assumption that during the evaporation process, the black hole remains classical till the end of its lifetime.

In Ref.~\cite{Dvali:2020wft}, it was demonstrated that Hawking's semi-classical calculations become inaccurate when the black hole has shed approximately half of its original mass. Hawking's approach overlooks the influence of emitted particles on the black hole, a neglect that becomes untenable as the energy of these particles approaches that of the black hole itself. Dvali et al. provided a pivotal realization that this influence results in a universal phenomenon termed {\it burden}, which effectively curtails further evaporation to a significant degree. At its core, the phenomenon of memory-burden~\cite{Dvali:2018xpy,Dvali:2018ytn,Dvali:2024hsb} suggests that when a system bears a substantial load of quantum information, it becomes stabilized. Consequently, for the system to decay, it must transfer the memory pattern from one set of modes to another. As indicated in Ref.~\cite{Alexandre:2024nuo}, this process becomes increasingly challenging as the system size expands. As a result, the quantum information retained in the memory acts back upon the system (formally called the backreaction), contributing to its stabilization, typically occurring no later than halfway through its decay. 

Motivated by this, in this study, we have delved into the potential for simultaneously generating observed DM abundance alongside baryon asymmetry (BAU) solely through PBH evaporation while considering the quantum effect of memory burden\footnote{In Ref.~\cite{Haque:2024eyh} the parameter space for DM from memory-affected PBH has been explored.}. While the DM could possess any intrinsic spin, we have adopted the standard approach of vanilla leptogenesis for generating baryon asymmetry. This involves the production of heavy right-handed neutrinos (RHN) during PBH evaporation, which subsequently undergoes CP-violating decay, leading to the observed matter-antimatter asymmetry. Finally, we discuss how such a framework might manifest in proposed gravitational wave (GW) detectors due to the presence of a detectable GW  spectrum induced by fluctuations in the PBH number density~\cite{Domenech:2020ssp, Papanikolaou:2020qtd, Balaji:2024hpu}. It is essential to highlight that a precise comprehension of the evaporation process beyond the semi-classical realm remains elusive. Consequently, our findings should be interpreted as a preliminary indication of how the constraints on PBHs evolve beyond the framework provided by Hawking's theory.

The paper is organized as follows. In Sec.~\ref{sec:pbh evaporation}, we briefly discuss the dynamics of PBH evolution with the memory effect included. We then explain in detail the intricacies of particle production from PBH evaporation in Sec.~\ref{sec:number}. The parameter space satisfying observed DM and baryon abundance is elaborated in Sec.~\ref{sec:cogenesis}. The effect of PBH evaporation on dark radiation is discussed in Sec.~\ref{sec:DR}. The observational aspect of the framework via induced GWs is discussed in Sec.~\ref{sec:gw}. Finally, we conclude in Sec.~\ref{sec:concl}.
\section{Effects of memory burden on PBH evaporation}
\label{sec:pbh evaporation}
Before delving into the details of the memory burden, we begin reviewing some necessary details on PBH dynamics. We then provide a brief discussion on the memory burden effect, especially highlighting its effect on PBH evaporation. We consider for simplicity that the PBH forms by the collapse of a sharply peaked primordial spectrum and,
therefore, the mass function of PBH is essentially monochromatic. Additionally, the PBHs are assumed to be of Schwarzschild type, i.e.,
without any spin and charge.
\subsection{Dynamics of PBH evaporation}
The mass of the PBH formed due to the gravitational collapse is closely related to the horizon size at the point of formation as~\cite{Fujita:2014hha,Masina:2020xhk}
\begin{align}\label{eq:min}    
\Min=\frac{4}{3}\,\pi\,\gamma\,\Bigl(\frac{1}{H\left(T_\text{in}\right)}\Bigr)^3\,\rho_\text{rad}\left(T_\text{in}\right)\,,
\end{align}
where $H(\Tin)$ corresponds to the Hubble rate during radiation domination at the time of PBH formation. Therefore, radiation temperature at the point of PBH formation
\begin{align}\label{eq:Tin}
& \Tin=\left(\frac{1440\,\gamma^2}{g_\star(\Tin)}\right)^{1/4}\,\left(\frac{M_P}{\Min}\right)^{1/2}\,M_P\,.    
\end{align}
Here, $\gamma$ represents the efficiency factor, which defines what fraction of the total mass inside the Hubble radius collapses to form PBHs. $g_\star(\Tin)$ is the relativistic degrees of freedom associated with the thermal bath at the point of formation. For the formation during standard domination $\gamma\sim 0.2$ \cite{Carr:1974nx}.
The time of the PBH formation is given as 
\begin{align}\label{eq:tin}
t_{\rm in}=\Min/(8\pi\gamma\Mpl^2)\,,  
\end{align}
where we have considered $H(t)=\frac{1}{2t}$ because of radiation dominated Universe.

Given a PBH with mass $\mbh$, the corresponding entropy, $S$, and Hawking temperature, $T_{\rm BH}$, reads
\begin{align} 
& T_{\rm BH}=\f{\Mpl^2}{\mbh}\,, & 
S(\mbh) =\f{1}{2}\l(\f{\mbh}{\Mpl}\r)^2=\f{1}{2}\l(\f{\Mpl}{T_{\rm BH}}\r)^2\,,
\label{eq:TbhS}
\end{align}
where $\Mpl= 1/\sqrt{8 \pi G_N}\simeq 2.4\times 10^{18}$ GeV
is the reduced Planck mass. Once PBH is formed, it can evaporate by emitting Hawking radiation~\cite{Hawking:1974rv, Hawking:1975vcx}. The mass loss rate for PBH can be parametrized as~\cite{MacGibbon:1991tj}
\bea 
\f{d\mbh}{dt}=-\epsilon\,\f{\Mpl^4}{\mbh^2}\,, \quad\text{with}\quad\epsilon =\f{27}{4}\f{\pi g_{\star,H}(T_\text{BH})}{480}\,,    
\label{eq:dmdt}
\eea 
where 
\begin{equation}
g_{\star,H}(T_\text{BH})\equiv\sum_i\omega_i\,g_{i,H}\,; g_{i,H}=
    \begin{cases}
        1.82
        &\text{for }s=0\,,\\
        1.0
        &\text{for }s=1/2\,,\\
        0.41
        &\text{for }s=1\,,\\
        0.05
        &\text{for }s=2\,,\\
    \end{cases}
    \label{eq:gstTBH}
\end{equation}
with $\omega_i=2\,s_i+1$ for massive particles of spin $s_i$, $\omega_i=2$ for massless species with $s_i>0$ and $\omega_i=1$ for $s_i=0$. At temperatures $T_\text{BH}\gg T_\text{EW}\simeq 160$ GeV, PBH evaporation emits the full SM particle spectrum according to their $g_{\star,H}$ weights, while at temperatures below the MeV scale, only photons and neutrinos are emitted. For $T_\text{BH}\gg 100$ GeV (corresponding to $M_\text{BH}\ll 10^{11}$ g), the particle content of the SM corresponds to $g_{\star,H}(T_\text{BH})\simeq 106.75$. The factor $27/4$ accounts for the graybody factor\footnote{Detailed expression for greybody factor can be found, for example, in Refs.~\cite{Auffinger:2020afu,Masina:2021zpu,Cheek:2021odj}.}. The negative sign on the right-hand side is due to the fact that the PBH mass decreases with time due to evaporation.

On integrating Eq.~\eqref{eq:dmdt} from the formation time $t_{\rm in}$, the PBH mass at any later time $t$ can be obtained
\bea 
\mbh(t)=\Min[1-\Gamma_{\rm BH}^0(t-t_{\rm in})]^{1/3}\,,
\label{eq:mbh1}
\eea 
where $\Min$ is given by Eq.~\eqref{eq:min}, and
\begin{align}
\Gamma_{\rm BH}^0=3\,\epsilon\,\Mpl^4/\Min^3\,,   
\end{align}
is the decay width associated with the PBH evaporation. The lifetime $t_{\rm ev}$ of the PBH can be obtained by substituting $\mbh(t_{\rm ev})=0$ in Eq.~\eqref{eq:mbh1},  
\bea 
t_{\rm ev}=\f{1}{\Gamma_{\rm BH}^0}=\f{\Min^3}{3\,\epsilon\, \Mpl^4}
\simeq 2.5\times 10^{-28} \left(\frac{\Min}{1~\rm{g}}\right)^3~\rm{s}\,,
\label{eq:tev1}
\eea 
where we supposed $t_{\rm ev} \gg \tin$. From Eq.(\ref{eq:tev1}), we recover that a PBH of mass $\gtrsim 10^{15}$ g should not have decayed yet, whereas PBH of mass below $\lesssim 10^9$ g should have decayed before the onset of BBN, which typically happens around time $t_{\rm BBN}\sim  1$ second. The initial PBH abundance is characterized by the dimensionless parameter $\beta$ that is defined as
\bea\label{eq:beta}
\beta\equiv\frac{\rho_\text{BH}\left(T_\text{in}\right)}{\rho_R\left(T_\text{in}\right)}\,,
\eea
that corresponds to the ratio of the initial PBH energy density to the SM energy density at the time of formation. It is worth noting that the evaporation process outlined above operates under semi-classical assumptions. This means, as a black hole evaporates, it gradually reduces in size while adhering to consistent semi-classical relationships between its key parameters, such as mass, radius, and temperature, in other words, such a semi-classical process is self-similar in nature. Eventually, this process culminates in a final burst as $\mbh\to 0$. However, the semi-classical approximation cannot hold
throughout the entirety of the black hole's lifetime.
\subsection{The effect of memory-burden}
For the purpose of our analysis, we consider that the semi-classical regime (Hawking's approach) is valid until 
\begin{align}
\mbh=q\Min\,.    
\end{align}
In Refs.~\cite{Alexandre:2024nuo,Thoss:2024hsr}, it has been 
proposed that the quantum
effects start to become important when $\mbh=\frac{1}{2}\Min$ or $q=\frac{1}{2}$. Since, because of evaporation, the PBH mass diminishes with time, the above relation implies $q<1$. We consider $t_q$ to be the time scale at which the semi-classical phase ends, beyond which the self-similar regime no longer remains valid. Then, from Eq.~\eqref{eq:mbh1} we obtain 
\bea 
t_q= \f{1-q^3}{\Gamma_{\rm BH}^0}\,.
\label{eq:tq}
\eea 
As evident, with the substitution of $q=0$, one recovers the full evaporation time $t_{\rm ev}$. Once the mass of the PBH reaches $q\,\Min$, the quantum memory effect starts dominating. This we consider to be the {\it second phase} of PBH evolution. In this regime, the evolution of the PBH mass, as given by Eq.~\eqref{eq:dmdt}, modifies to\footnote{Note that $\mbh$ being almost constant during the evaporation process, so during the memory burden phase $S(\mbh)\sim S(q\,\Min)$, which corresponds to the approximation made in Ref.~\cite{Thoss:2024hsr}.}
\bea \label{Eq:memory}
\f{d\mbh}{dt}=-\f{\epsilon }{\l[S(\mbh)\r]^k}\f{\Mpl^4}{\mbh^2},
\label{eq:dmbhdt}
\eea 
where $S(\mbh)$ is defined in Eq.~\eqref{eq:TbhS}. We emphasize that this is just a parametrization to capture the effect of memory burden on PBH evaporation \footnote{It is important to note that Eq.~\eqref{Eq:memory} is based on assumptions derived from well-constructed toy models, that capture key properties of black hole dynamics, such as entropy production and breakdown of semiclassical approximation, etc., as outlined by Dvali et al. in Refs.~\cite{Dvali:2011aa,Dvali:2012en}. However, the present analysis does not account for the potential onset of classical instabilities that could lead to a faster decay of PBHs that hinder prolonged lifetime due to the memory burden effect.
}. There are no such inherent constraints on $k$. Since we understand that the backreaction effect slows the decay due to the excess entropy production surrounding the PBH, we consider $k$ a positive value. Thus, the exponent determines the efficiency of the backreaction effect. Throughout our analysis, we choose $k$ as a free parameter. Integrating Eq.~\eqref{eq:dmbhdt} we obtain
\bea 
\mbh=q\,\Min\l[1-\Gamma_{\rm BH}^k(t-t_q)\r]^{\frac{1}{3+2k}}\,,
\label{eq:mbh2}
\eea 
with
\begin{align}\label{eq:GammaBHk}
\Gamma_{\rm BH}^k= {2^k\,(3+2k)\,\epsilon}\,\Mpl
\l(\f{\Mpl}{q\Min}\r)^{3+2k}\,.    
\end{align}
From Eq.~\eqref{eq:mbh2}, we find that the second phase of evaporation takes place over a time $\sim 1/\Gamma_{\rm BH}^k$. The minimum allowed mass is set by the maximum energy of inflation, calculated assuming de-Sitter-like inflation by taking the restriction on the tensor to scale ratio $r<0.036$ from Planck together with the latest BICEP/$Keck$ data~\cite{Planck:2018jri,BICEP:2021xfz,Tristram:2021tvh}. The minimum allowed mass is roughly around $\sim 0.5$ g, which sets $k\lesssim3$ for complete evaporation before BBN.
\begin{figure}[t]
    \centering
    \includegraphics[scale=.37]{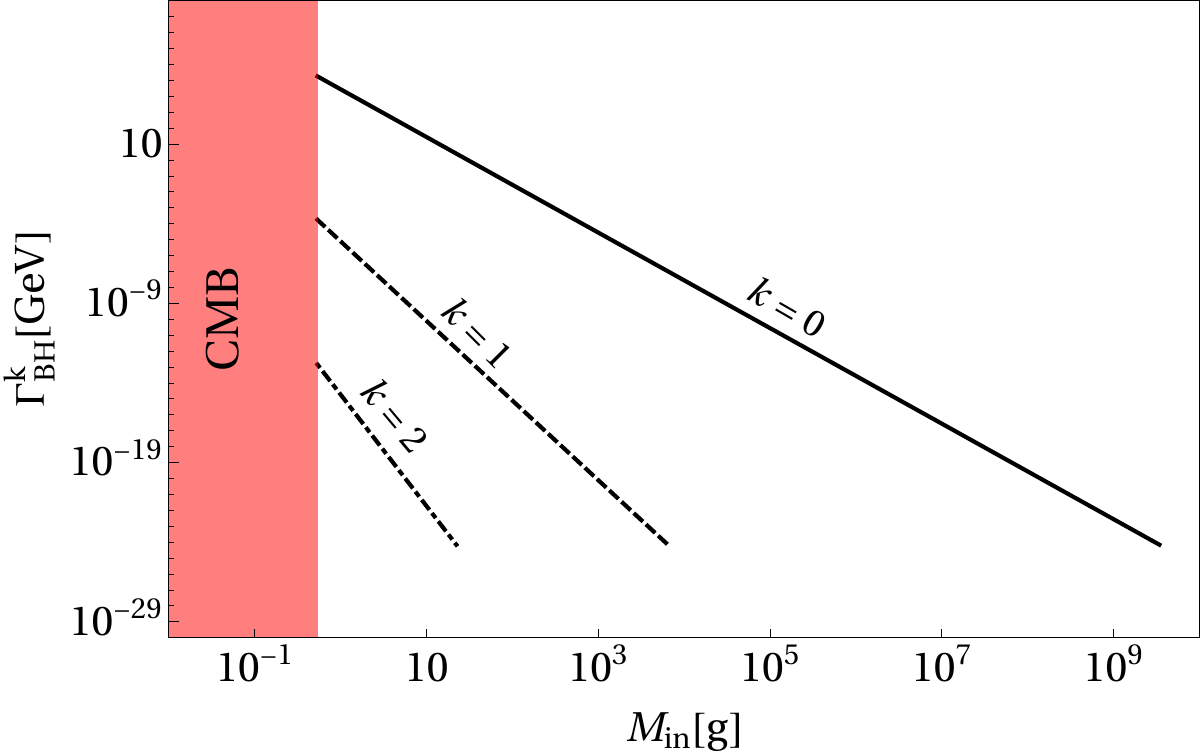}~~\includegraphics[scale=.37]{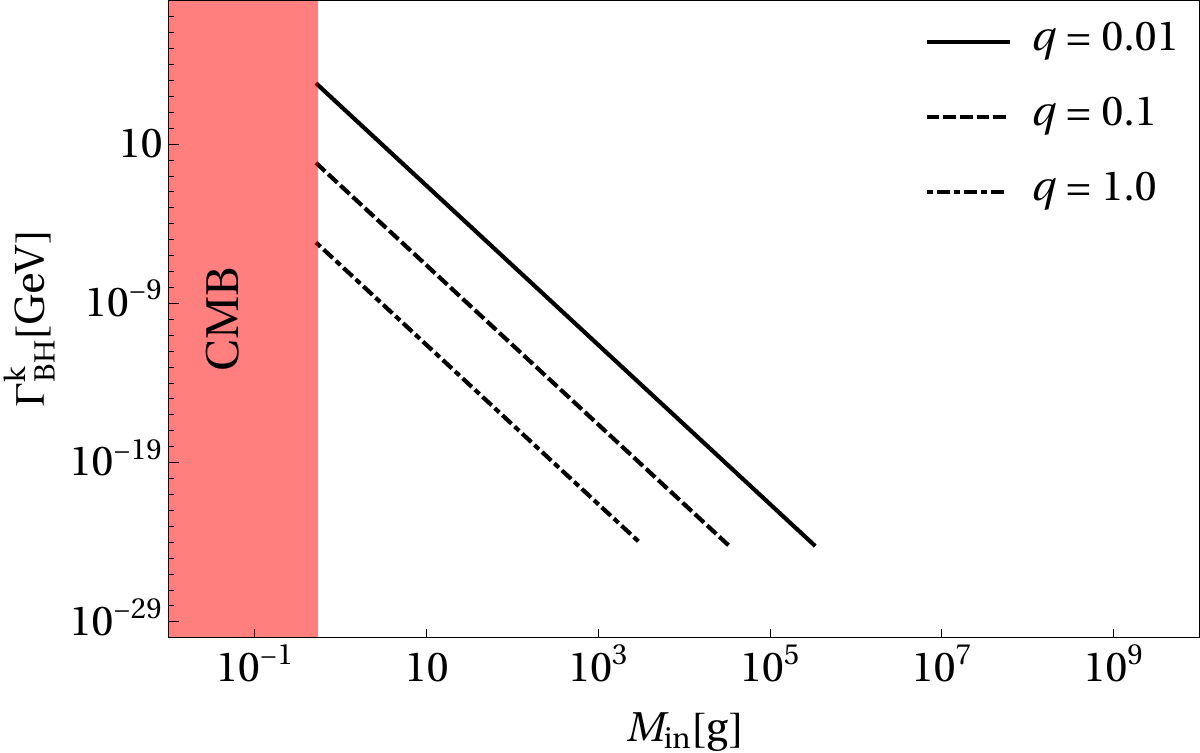}
    \caption{PBH decay width as a function of $\Min$, following Eq.~\eqref{eq:mbh2}. In the left panel, we have chosen different values of $k$ while fixing $q=0.5$, while in the right panel, different values of $q$ are considered for a fixed $k=1$. The shaded region is discarded from CMB, which is calculated considering the maximum energy scale of inflation. All curves satisfy the BBN bound.}
    \label{fig:gammak}
\end{figure}

\begin{figure}[t]
    \centering    \includegraphics[scale=.37]{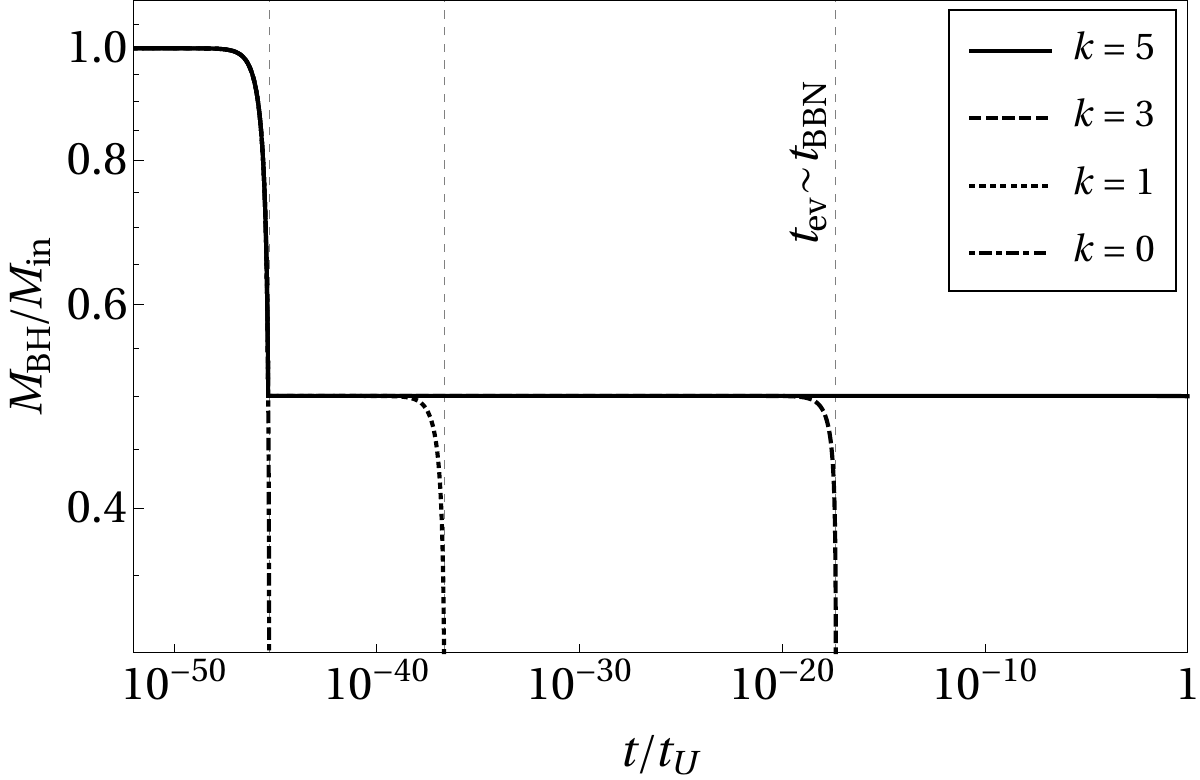}~~\includegraphics[scale=.37]{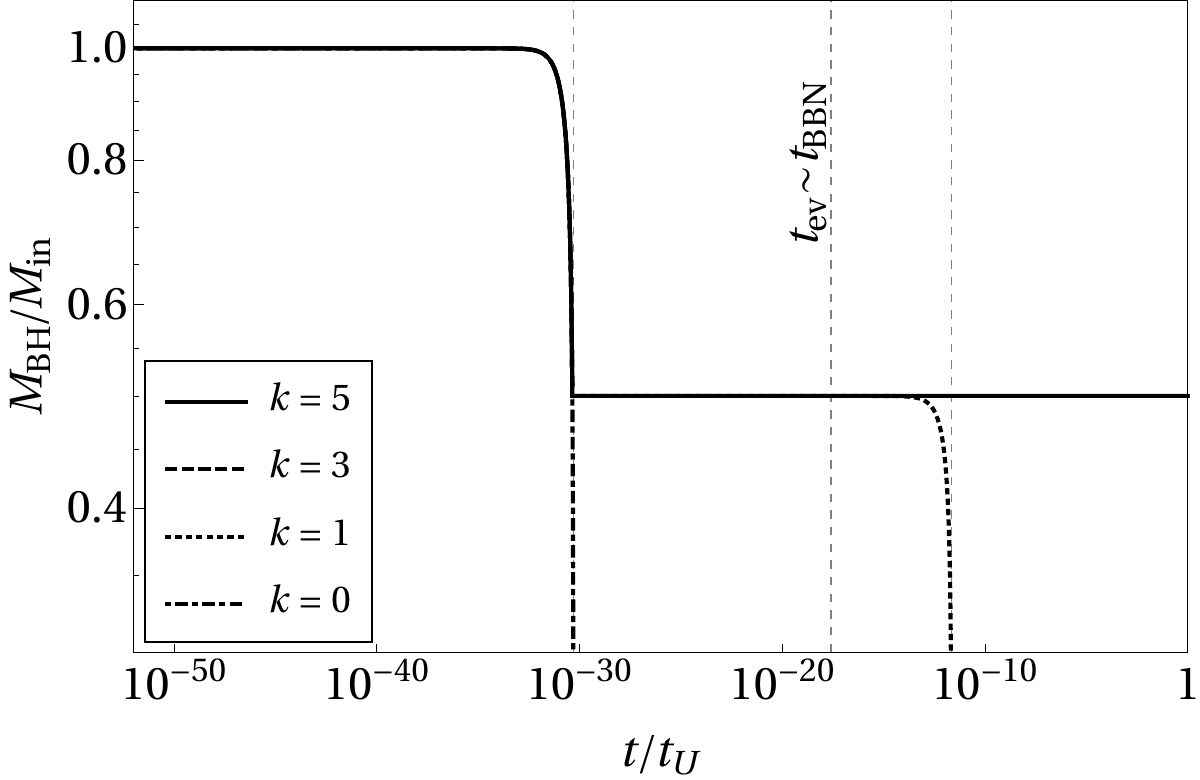}
    \caption{Evolution of PBH mass as a function of time for different choices of $k$. Here, $t_U\sim 4\times 10^{17}$ s in the horizontal axis correspond to the universe's current age. We choose $\Min=1$ g in the left panel and $\Min=10^5$ g in the right panel while fixing $q=0.5$ in either case. The vertical dashed lines indicate evaporation time corresponding to different choices of $k$.}
    \label{fig:Min-k}
\end{figure}

In Fig.~\ref{fig:gammak} we show the variation of $\Gamma_{\rm BH}^k$ with $\Min$. As one can see from the left panel, for a fixed $q$, a larger $k$ results in late evaporation of the PBH since the total evaporation
time is given by
\begin{align}\label{eq:tevTot}
t_{\rm ev}^k=t_q+1/\Gamma_{\rm BH}^k\simeq 1/\gammabh^k\,.    
\end{align}
Also note, for different $k$-values, the curves have different end-points such that the BBN bound is satisfied, i.e., $t_{\rm ev}\lesssim 1$ sec. Since a larger $k$ results in longer PBH-lifetime, hence corresponding constraint on PBH mass from BBN is stronger. On the other hand, the dependence on $q$ is rather mild for a fixed $k$ as one can notice from the right panel of Fig.~\ref{fig:gammak}. The bottom line is, breakdown of the self-similar regime results in a suppressed PBH decay rate. The evolution of PBH mass as a function of cosmological time $t$ is shown in Fig.~\ref{fig:Min-k}, for different choices of $k$ and a fixed $q=0.5$. In the left panel we have considered $\Min=1$ g. Following the conclusions drawn from Fig.~\ref{fig:gammak}, here as well we find that a larger $k$ results in a longer PBH lifetime. This effect becomes more prominent for heavier PBHs, as we see from the right panel, where $\Min=10^5$ g is considered. The vertical dashed lines in each figure denotes the PBH evaporation time, showing, for $k>3$ and $q=0.5$, PBH of mass 1 g remains stable even after BBN, while for $\Min=10^5$ g, this is true for $k>1$. 

\section{Particle production from memory-burdened PBH}
\label{sec:number}
As advocated in the last section, the memory burden results in a complete breakdown of the semi-classical treatment. If the breakdown happens at $t_q$, implying for $M_{\rm BH}>qM_{\rm in}$ there exists the semi-classical evaporation phase, which we will now refer as {\bf ph-I} (phase-I). Whereas for $M_{\rm BH}<qM_{\rm in}$  we have a memory burden phase where the semi-classical approximation is no longer valid, we refer to this phase as {\bf ph-II} (phase-II). In order to find the viable parameter space giving rise to observed DM relic abundance and correct baryon asymmetry, we first need to calculate the number of any massive species emitted during the process of PBH evanescence, which can further be utilized for the calculation of relevant relic abundances.  

Consider $N_j$ be the number of particle species $j$ with mass $m_j$ and intrinsic spin $g_j$, that is emitted from the evaporation of a single PBH. We divide $N_j$ into two parts: (i) $N_{1j}$ denotes the number of particles emitted in {\bf ph-I} and (ii) $N_{2j}$ is the number of particles emitted in {\bf ph-II}. The emission rate of particle $j$ with mass $m_j$ and internal degrees of freedom $g_j$, per unit energy per unit time, due to Hawking radiation in {\bf ph-I} is given by~\cite{Perez-Gonzalez:2020vnz,Cheek:2021cfe,Cheek:2021odj,Haque:2024cdh,RiajulHaque:2023cqe}
\bea 
\f{d^2N_{1j}}{dE\,dt} &=& \f{27}{4}\,\pi\,r_S^2\,\f{g_j}{2\pi^2} \f{E^2}{\exp(E/\Tbh)\pm 1} = \f{27}{4}\,\f{g_j}{32\,\pi^3}\f{(E/\Tbh)^2}{\exp(E/\Tbh)\pm 1}\,,
\label{eq:d2Nj1}
\eea 
where the Schwarzchild radius is given by $r_S=\mbh/(4\pi\Mpl^2)$. 
The sign $\pm$ is used for fermionic and bosonic particles, respectively. Integrating over Eq.~\eqref{eq:d2Nj1} we find
\bea 
\f{dN_{1j}}{dt} =\f{27}{4} \f{\xi\, g_j\,\zeta(3)}{16\,\pi^3} \f{\Mpl^2}{\mbh}, \quad \text{where}
\quad \xi=\begin{cases}
    1& \text{for bosons},\\
    \f{3}{4}& \text{for fermions}
\end{cases}\,,
\label{eq:dNj1}
\eea 
which is the rate of emission of particles from a given PBH of mass $\mbh$. In the {\bf ph-II}, because of the suppressed evaporation rate, the emission rate is parametrized as
\bea\label{eq:d2Nj2} 
\f{d^2N_{2j}}{dE\,dt} = \f{1}{\l[S(\mbh)\r]^k}\f{d^2N_{1j}}{dE\,dt}\,, 
\eea 
that leads to
\bea\label{eq:dNj2}
\f{dN_{2j}}{dt} =\f{27}{4} \f{\xi g_j\,\zeta(3)\,2^k}{16\,\pi^3} \f{\Mpl^{2+2k}}{\mbh^{1+2k}}\,.
\eea 

Depending on the mass of the emitted particle relative to the PBH temperature at the point of formation, one should further distinguish two cases:
(a) if $m_j<\Tbh^{\rm in}$, the production of particles happens throughout the lifetime of the PBH, i.e., from the initial time $t_{\rm in}$ to a final time $t$, and (b) for $m_j>\Tbh^{\rm in}$, the evaporation begins from a time $t_j$ corresponding to which $m_j=\Tbh(t_j)$, 
\bea 
t_j\simeq
\begin{cases}
t_{\rm ev}\l[1-\f{\Mpl^6}{\Min^3 m_j^3} \r]& \text{for}~t_j<t_q,\\
&\\
t_{\rm ev}^{k}\l[1-\l(\f{\Mpl^2}{q\Min m_j}\r)^{3+2k} \r]& \text{for}~t_j>t_q\,,
\end{cases}
\label{eq:tj}
\eea  
where we have utilized Eq.~\eqref{eq:mbh1} and Eq.~\eqref{eq:mbh2}. 

Let us now first consider the case $m_j<T_{_{\rm BH}}^{\rm in}$. In order to determine $N_{1j}$ produced during {\bf ph-I}, we exploit Eq.~\eqref{eq:mbh1} and integrate Eq.~\eqref{eq:dNj1} in the limit $t_{\rm in}\leq t\leq t_q$, which gives rise to
\begin{align}\label{eq:N1ja}
& N_{1j}=\frac{27}{128}\,\frac{\xi\,g_j\,\zeta(3)}{\pi^3}\,\frac{(1-q^2)}{2\,\epsilon}\,\frac{\Min^2}{M_P^2}\,, 
\end{align}
where we have ignored a term proportional to $\tin$ as $\tin\ll t_{\rm q}$. To compute the total number $N_{2j}$ during {\bf ph-II}, we integrate Eq.~\eqref{eq:dNj2} in the range $\left[t_q,\,t\right]$, 
\begin{align}\label{eq:N2ja}
& N_{2j}=\frac{27}{128}\,\frac{\xi\,g_j\,\zeta(3)}{\pi^3\,\epsilon}\,q^2\,\frac{\Min^2}{M_P^2}\,\left[1-\left(1-\frac{t-t_q}{t_{\rm ev}^k}\right)^\frac{2}{3+2\,k}\right]\,,   
\end{align}
where we have utilized Eq.~\eqref{eq:mbh2}. Note that the above expression simply vanishes for $q=0$, implying the evaporation solely through only Hawking evaporation. For $m_j<T_{_{\rm BH}}^{\rm in}$, the total number of particles emitted by one PBH at the end of its lifetime is given by
\begin{align}
& N_j\Big|_{m_j<T_{_{\rm BH}}^{\rm in}}=N_{1j}+N_{2j}=\frac{27}{128}\,\frac{\xi\,g_j\,\zeta(3)}{\pi^3\,\epsilon}\,\frac{\Min^2}{M_P^2}\,\left[1-q^2\,\left(\frac{t_q}{t_{\rm ev}^k}\right)^\frac{2}{3+2\,k}\right]
\nonumber\\&
\simeq \frac{27}{128}\,\frac{\xi\,g_j\,\zeta(3)}{\pi^3\,\epsilon}\,\frac{\Min^2}{M_P^2}\,, 
\end{align}
where $t_q/t_{\rm ev}^k\ll1$ is 
implied. 

In the reverse case, i.e., $m_j>T_{_{\rm BH}}^{\rm in}$, we have two sub-cases corresponding to Eq.~\eqref{eq:tj}. For $t_j<t_q$ we thus obtain $N_{1j}$ by integrating Eq.~\eqref{eq:dNj1} between $[t_j,t_q]$, we have
\begin{align}
& N_{1j}=\frac{27}{128}\,\frac{\xi\,g_j\,\zeta(3)}{\pi^3\,\epsilon}\,\left[\left(\frac{M_P^2}{m_j^2}\right)^2-\left(\frac{q^2\,\Min^2}{M_P^2}\right)\right]\,,    
\end{align}
while $N_{2j}$ remains same as that obtained in Eq.~\eqref{eq:N2ja}. For $t_j>t_q$, there is no production of particles during {\bf ph-I}, implying $N_{1j}=0$, whereas
\begin{align}
& N_{2j}=\frac{27}{128}\,\frac{\xi\,g_j\,\zeta(3)}{\pi^3\,\epsilon}\,\frac{q^2\Min^2}{M_P^2} 
\left[\left(1-\f{t_j-t_q}{t_{\rm ev}}\right)^{\frac{2}{3+2k}} 
-\left(1-\frac{t-t_q}{t_{\rm ev}}\right)^{\frac{2}{3+2k}}\right],    
\end{align}
obtained by integrating within the range $[t_j,\,t]$. In this case, setting $t=t_{\rm ev}^k>t_j>>t_q$ and utilizing Eq.(\ref{eq:tj}), the total number of particles emitted from a single BH can be expressed as 
\begin{align}
& N_j\Big|_{m_j>T_{_{\rm BH}}^{\rm in}}
\simeq \frac{27}{128}\,\frac{\xi\,g_j\,\zeta(3)}{\pi^3\,\epsilon}\,\left(\frac{M_P}{m_j}\right)^2\,,
\end{align}
which will be the same in the case of $t_j<t_q$.
We thus find,
\begin{align}
& N_j\simeq\frac{27}{128}\,\frac{\xi\,g_j\,\zeta(3)}{\pi^3\,\epsilon}
\begin{cases}
\left(\frac{\Min}{M_P}\right)^2\,, & \text{for}~ m_j<T_{_{\rm BH}}^{\rm in},    
\\[10pt] 
\left(\frac{M_P}{m_j}\right)^2\,, & \text{for}~m_j>T_{_{\rm BH}}^{\rm in}\,.
\end{cases}
\end{align}
It is crucial to note here that the above expression does not contain any information about the backreaction, which is expected since we are interested in production from complete evaporation. Once PBH completes its evaporation, the number of particles emitted from a PBH only depends on the mass of the BH and emitting particle; the underlying process is not important here. 
\begin{figure}
    \centering
    \includegraphics[scale=.5]{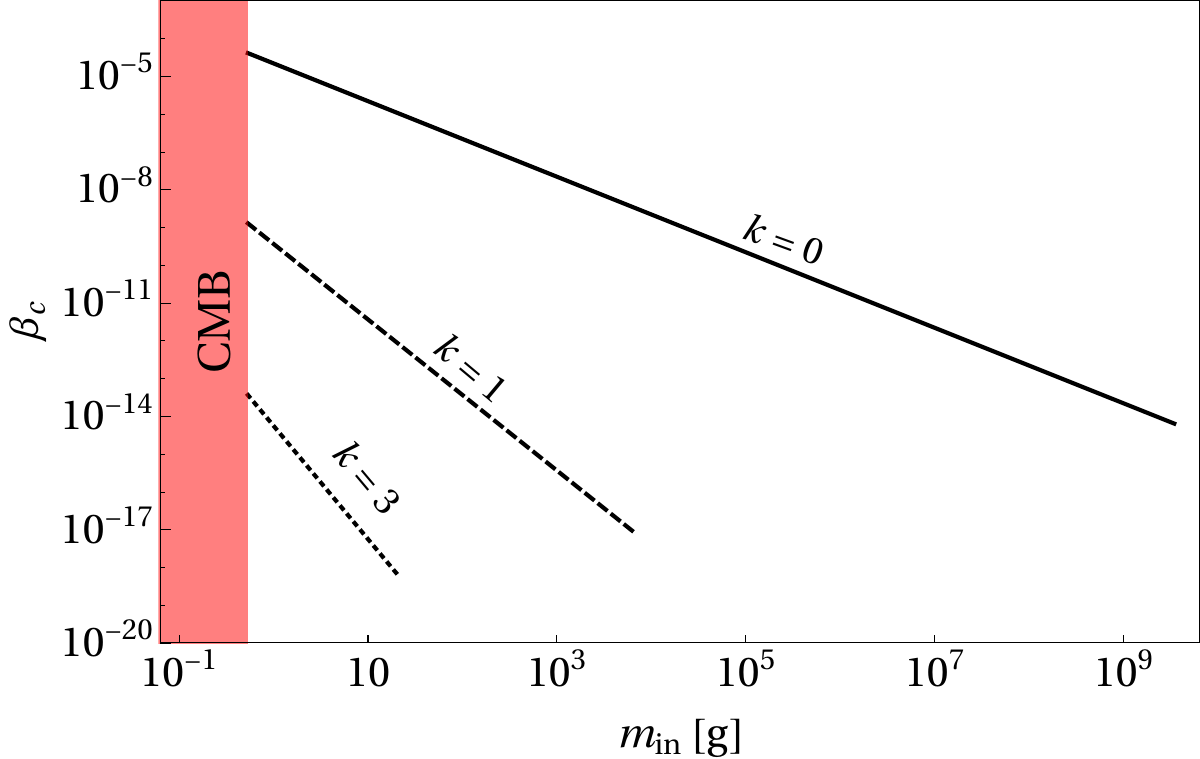}
    \caption{$\beta_c$ as function of $\Min$, following Eq.~\eqref{eq:betac}, where different lines correspond different choices of $k$. Here we have chosen $q=0.5$. The shaded region is disallowed by CMB bound on the scale of inflation. All curves satisfy BBN bounds.}
    \label{fig:betac}
\end{figure}

Since the PBH energy density scales like non-relativistic matter $\sim a^{-3}$, while the radiation energy density scales as $\rho_R\sim a^{-4}$, therefore, an initially radiation-dominated universe will eventually become matter-dominated if the PBHs have not evaporated by then. 
There exists a critical $\beta$ value, above which PBHs come to dominate the energy content of the universe at some time $t_{\rm eq}^{\rm BH}$. This scenario corresponds to the so-called PBH domination. Utilizing Eq.~\eqref{eq:beta}, one can obtain the expression for critical $\beta$-value $\beta_{\rm c}$ as~\cite{Masina:2020xhk}
\bea\label{eq:betac}
\beta_{\rm c}=\sqrt{\frac{t_{\rm ev}^k}{t_{\rm in}}}= \l(\f{\Mpl}{q\,\Min}\r)^{1+k} \sqrt{\f{(3+2k)2^k\epsilon }{8\,q^3\pi\gamma}}\,.
\eea 
In Fig.~\ref{fig:betac}, we show how $\beta_c$ varies with $\Min$ for different choices of $k$. Since increasing the $k$ value of the memory burden parameter delays the evaporation process, for a given $\Min$, a larger $k$ results in smaller $\beta_c$.


For $\beta>\beta_{\rm c}$, there will be a PBH domination before PBH decays. Therefore, 
\bea
&&
H (\aev)^2=\frac{\rho_{\rm rad}(\aev)}{3M_P^2}
\equiv\left(\frac{2\,\Gamma_{\rm BH}^k}{3}\right)^2\,,
\eea
which results in an evaporation temperature,
\begin{align}\label{eq:Tev1}
& T_{\rm ev}=\Mpl\,\l(\f{40}{\pi^2\,g_*(\Tev)}\r)^{1/4}\,\l[2^k\,(3+2k)\,\epsilon\l(\f{\Mpl}{q\,\Min}\r)^{3+2k}\r]^{1/2}\,,
\end{align}
where we have used Eq.~\eqref{eq:GammaBHk} and the fact that in a matter (PBH)-dominated universe, $H(t_{\rm ev}^k)=2/\left(3\,t_{\rm ev}^k\right)$. $g_*(\Tev)$ indicates the relativistic degrees of freedom associated with the thermal bath calculated at the end of evaporation. Evidently, for a given $k$ and $q$, the evaporation temperature decreases with the increase in $\Min$. Since we can grossly estimate that the energy density in BHs goes into radiation after evanescence, therefore at $t=t_{\rm ev}^k$,
\begin{align}
& \frac{\rho_{\rm BH}}{3\,M_P^2}=\left(\frac{2\,\Gamma_{\rm BH}^k}{3}\right)^2\,,    
\end{align}
which leads to the PBH number density at the evaporation point
\bea
n_{_{\rm BH}} (a_{\rm ev})=\f{4}{3}\, \Mpl^3\,(3+2k)^2\,2^{2k}\,\epsilon^{2}\, \l(\f{\Mpl}{q\,\Min}\r)^{7+4k}\,,
\label{Eq:nBH1}
\eea
where Eq.~\eqref{eq:mbh2} has been used.

\subsection*{Gravity-mediated particle production from bath}
\label{sec:grav}
Pure gravitational production of particles is also feasible via 2-to-2 scattering of the bath particles via $s$-channel mediation of massless graviton. The interaction rate density for such a process reads~\cite{Garny:2015sjg, Tang:2017hvq,Garny:2017kha, Bernal:2018qlk,Barman:2021ugy,Barman:2021qds,Haque:2021mab}
\begin{equation}
    \gamma(T) = \zeta\, \frac{T^8}{M_P^4}\,,
\end{equation}
with $\zeta\simeq 1.9\times 10^{-4}$ (real scalar), $\zeta\simeq 1.1\times 10^{-3}$ (Dirac fermion) or $\zeta\simeq 2.3\times 10^{-3}$ (vector boson). Such production is inevitable because of the democratic nature of gravitational coupling. The Boltzmann equation, therefore, reads
\begin{equation}\label{eq:beq-UV}
\dot n_j + 3\,H\,n_j=\gamma(T)\,.    
\end{equation}
For temperatures much lower than the reheat temperature i.e., $T\ll\Trh$, the yield $Y\equiv n_j/s$ can be analytically obtained by integrating Eq.~\eqref{eq:beq-UV},
\begin{equation}\label{eq:grav-yield1}
Y_0 = \frac{45\,\zeta}{2\,\pi^3\,g_{\star s}}\,\sqrt{\frac{10}{g_\star}}\,\left(\frac{T_\text{RH}}{M_P}\right)^3\,,    
\end{equation}
considering $m_j\ll\Trh$. On the other hand, in the case where DM is heavier than the reheating temperature (but still
lighter than the highest temperature reached by the SM thermal bath~\cite{Giudice:2000ex}), then the DM is produced during the reheating era, with a yield
\begin{equation}\label{eq:grav-yield2}
Y_0 =  \frac{45\,\zeta}{2\,\pi^3\,g_{\star s}}\,\sqrt{\frac{10}{g_\star}}\,\frac{\Trh^7}{M_P^3\,m_j^4}\,.     
\end{equation}
Now, the particle produced via gravitational UV freeze-in shall undergo dilution due to evaporation of the PBH, that can be quantified as~\cite{Bernal:2021yyb,Bernal:2022oha} 
\begin{align}
& \frac{S(T_\text{in})}{S(T_\text{ev})}\simeq
\begin{cases}
1\,,& \beta < \beta_c\\[10pt]
T_\text{ev}/T_\text{eq}^{\rm BH}\,, & \beta>\beta_c\,,
\end{cases}
\end{align}
where we define $S=a^3\,s(T)$, $\Tev$ is given by Eq.~\eqref{eq:Tev1}, and $T_\text{eq}^{\rm BH}$ is defined as the epoch of equality between SM radiation and the PBH energy densities $\rho_R(T_\text{eq}^{\rm BH})=\rho_\text{BH}(T_\text{eq}^{\rm BH})$, and is given by
\begin{equation}
T_\text{eq}^{\rm BH}=\beta\,T_\text{in}\,\left(\frac{g_\star(\Tin)}{g_\star(\Teq)}\right)\,\left(\frac{\gss(\Teq)}{\gss(\Tin)}\right)\simeq\beta\,\Tin\,\,
\end{equation}
where $\gss(T_{\rm eq}^{\rm BH})$ and $g_\star(T_{\rm eq}^{\rm BH})$ are the relativistic DoFs associated with entropy and thermal bath, respectively calculated at the point of early radiation-matter equality. 
Following~\cite{Barman:2021ost,Bernal:2022oha}, here we will consider $\Trh=\Tin$ in order to perform a conservative analysis, where $\Tin$ is given by Eq.~\eqref{eq:Tin}. The right relic abundance is satisfied via
\begin{align}
& \mdm\times Y_0\times \frac{S(T_\text{in})}{S(T_\text{ev})}=\Omega_{\rm DM}\,h^2\frac{1}{s_0}\,\frac{\rho_c}{h^2}=4.3\times 10^{-10}\,{\rm GeV}\,,    
\end{align}
with $s_0$ being the comoving entropy density at the present epoch and $\rho_c$ being the critical energy density of the Universe. Note that since we assume that the  PBHs are formed during a radiation-dominated era, $T_{\rm RH}\geq \Tin$. 
\section{Cogenesis from memory-burdened PBH}
\label{sec:cogenesis}
In this section, we discuss the possibility of simultaneously satisfying the observed relic abundance of dark matter and correct baryon asymmetry purely from PBH evaporation, considering the effect of memory burden. We concentrate on the scenario of PBH dominance before it evaporates, which means $\beta>\beta_c$, such that the effect of PBH on the allowed parameter space becomes more pronounced.
\subsection{Production of dark matter}
We consider the production of dark matter (DM) purely from PBH evaporation, where the corresponding relic abundance today reads~\cite{Fujita:2014hha,RiajulHaque:2023cqe} 
\bea
\Omega_{\rm DM}\,h^2= 1.6\times 10^8\,\frac{g_0}{g_{\rm ev}}\frac{ n_{\rm DM}(\aev)}{T_{\rm ev}^3}\,\frac{\mdm}{\text{GeV}}\,,
\label{eq:dmrel}  
\eea
where $n_{\rm DM}(\aev)$ is the number density of DM at the end of evaporation. Here $g_{\rm ev}\simeq 106.7$ and $g_0\simeq 3.91$ are the effective numbers of light species for entropy at the end of evaporation and present-day, respectively. For complete evaporation, we can write
\begin{align}
n_{\rm DM}(\aev)=N_{\rm DM}\,n_{\rm BH}(\aev)\,,    
\end{align}
where $N_{\rm DM}$ is the total number of DM particles emitted from the evaporation of a single PBH and $n_{\rm BH}(\aev)$ is the PBH energy density at the point of evaporation. Since we are interested in the PBH-domination scenario, hence, using Eq.~\eqref{Eq:nBH1}, together with Eq.~\eqref{eq:Tev1}, we obtain
\begin{align}
& \f{\Omega_{\rm DM}\,h^2}{0.12}=\xi\,g_{\rm DM}\,\l[2^k\,(3+2k)\r]^{1/2}\times
\begin{cases}
2.85\times10^6\times\frac{1}{q^2}\, \l(\f{\Mpl}{q\,\Min}\r)^{\frac{2k+1}{2}} 
\,\frac{\mdm}{\text{GeV}}  & \text{for}~\mdm<T_{_{\rm BH}}^{\rm in},
\\[10pt]
1.64\times10^{43}\times\l(\f{\Mpl}{q\,\Min}\r)^{\frac{2k+5}{2}} 
\,\frac{\text{GeV}}{\mdm}  & \text{for}~\mdm>T_{_{\rm BH}}^{\rm in}\,,
\end{cases}
\end{align}
where $g_{\rm DM}$ is the intrinsic DoF for corresponding DM particle.
\begin{figure}
    \centering
   \includegraphics[scale=.37]{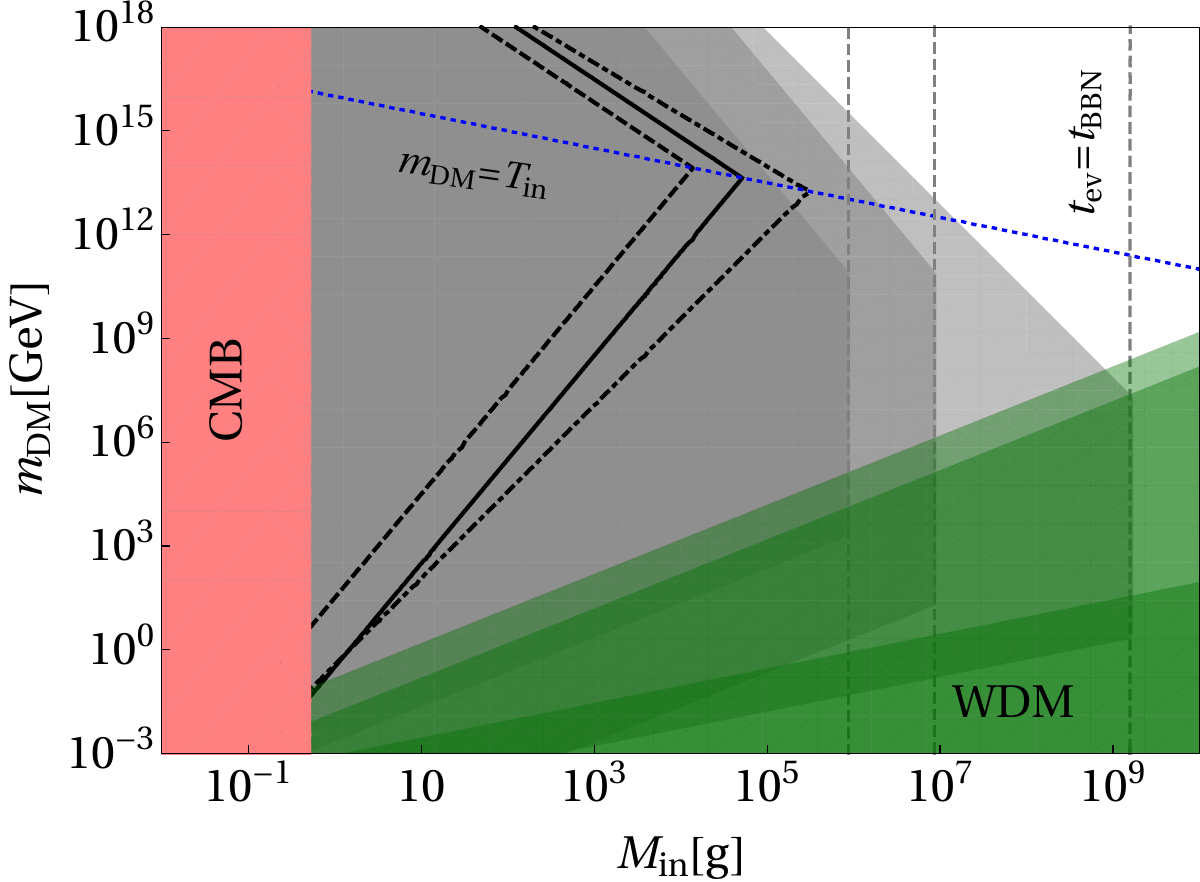}~~\includegraphics[scale=.37]{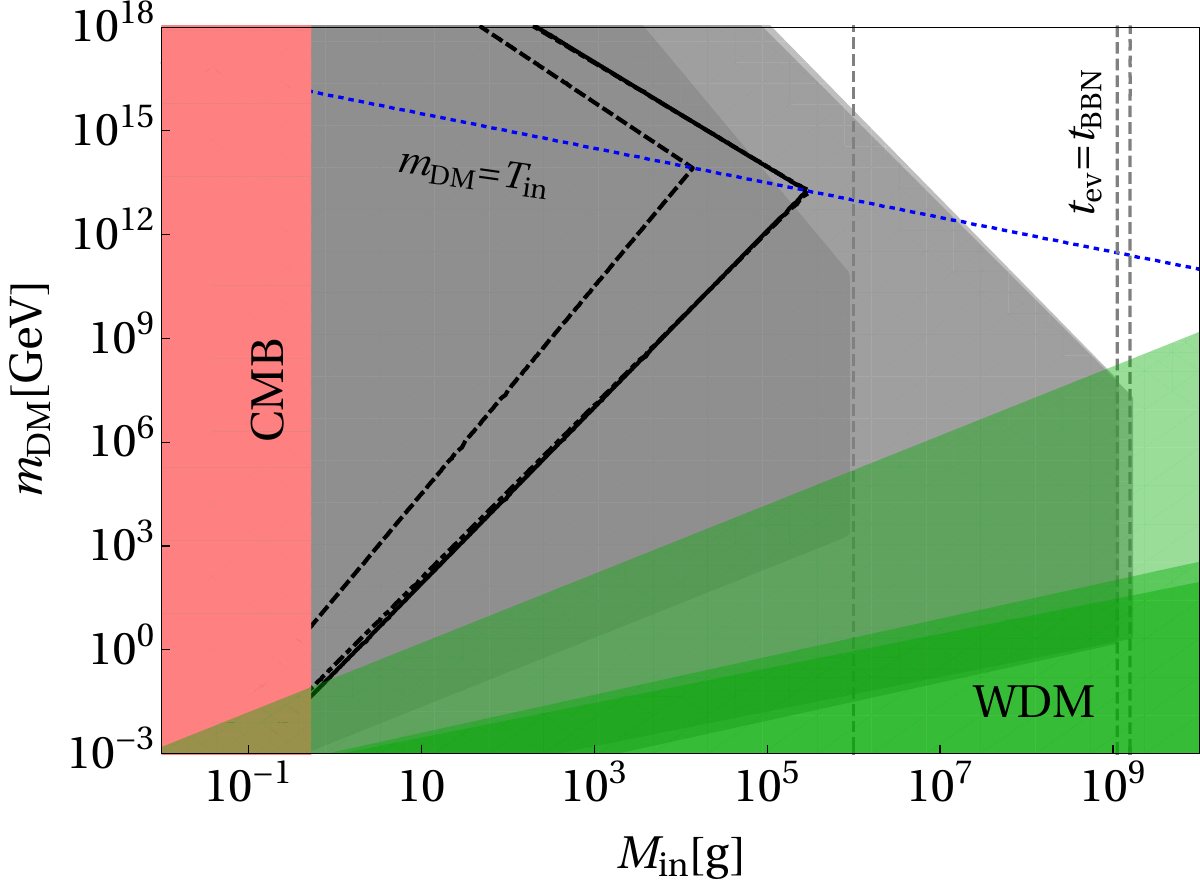}
   \caption{The gray shaded region corresponds to scalar DM overdensity for $\{k,\,q\}=\{0.5,\,0.5\};\,\{0.5,\,0.05\};\,\{0,\,1\}$ (left panel) and $\{k,\,q\}=\{0.5,\,0.5\};\,\{0.05,\,0.5\};\,\{0,\,1\}$ (right panel), from darker to lighter considering $\beta>\beta_c$. The pink-shaded region is disallowed from CMB constraint on the scale of the inflation. We show constraint from Ly-$\alpha$ on warm DM via the green-shaded region for each case. The black contours show correct DM abundance produced from a gravity-mediated process for $\{k,\,q\}=\{0.5,\,0.05\}\,;\{0.5,\,0.5\}\,;\{0,\,1\}$ shown via solid, dashed and dot-dashed curves respectively in the left panel and for $\{k,\,q\}=\{0.05,\,0.5\}\,;\{0.5,\,0.5\}\,;\{0,\,1\}$ shown via solid, dashed and dot-dashed curves respectively in the right panel.}
    \label{fig:relic}
\end{figure}

If the DM  is going to provide the full contribution to DM relic abundance, one has to ensure that it is cold enough in order not to disrupt the formation of large-scale structures. DM candidate, which is part of the thermal bath or produced from the thermal bath, should have a mass above a few keV in order to give required free-streaming of DM as constrained from Lyman (Ly)-$\alpha$ flux-power spectra~\cite{Irsic:2017ixq,Ballesteros:2020adh}. Such light DM mass is significantly constrained by the free
streaming of the ultra-relativistically-emitted DM particles. Following Refs~\cite{Fujii:2002jw,Masina:2020xhk}, one can derive a lower bound on the DM mass originating from PBH evaporation as
\begin{align}\label{eq:wdm}
& \mdm\gtrsim 10^4\,\langle E_{\rm DM}(t_{\rm eq})\rangle\,,    
\end{align}
where $\langle...\rangle$ stands for the average over PBH temperature. The average kinetic energy (KE) of the DM particles then reads
\begin{align}\label{eq:EDM}
\langle E_{\rm DM}(t_{\rm eq})\approx\langle E_{\rm DM}(t_{\rm ev})\rangle\,\frac{a_{\rm ev}}{a_{\rm eq}}\simeq \delta\,\Tbh\,\frac{T_{\rm eq}}{T_{\rm ev}}\,\left[\frac{\gss(T_{\rm eq})}{\gss(T_{\rm ev})}\right]^{1/3}\,,    
\end{align}
with $T_{\rm eq}\simeq 0.75$ eV being the temperature at the epoch of late radiation-matter equality (RME), and $\Tbh\simeq 1/(q\,\Min)$. Here $\gss(T_{\rm eq})$ and $\gss(T_{\rm ev})$ are the relativistic DoFs associated with entropy at the point of late radiation-matter equality and evaporation, respectively. To be conservative, we define the parameter $\delta\simeq 1.3$ such that $\langle E_{\rm DM}(t_{\rm ev})\rangle\approx\delta\,\Tbh$, following Refs.~\cite{Lennon:2017tqq,Masina:2020xhk}. Also, we assume entropy conservation from evaporation to RME. From Eq.~\eqref{eq:wdm}, one then finds 
\begin{align}
\mdm & \gtrsim 2.2\times 10^{-5}\,{\rm GeV}\,\left(\frac{q\,\Min}{M_P}\right)^\frac{2k+1}{2}\,\frac{1}{\sqrt{2^k\,\epsilon\,(2\,k+3)}}\,\left[\frac{\gss(T_{\rm eq})}{\gss(T_{\rm ev})}\right]^{1/3}
\nonumber\\&
\gtrsim 1.6\,\text{GeV}\left(\frac{\Min}{10\,\text{g}}\right)\,\left(\frac{q}{0.5}\right)\,,    
\end{align}
where in the last line, we have used $k=0.5$. Note that, for a given PBH mass, this bound becomes stronger with larger $k$ and $q$, as in that case, the PBH evaporation is delayed; as a result, the average kinetic energy of the radiated particles increases at RME.

In Fig.~\ref{fig:relic}, we illustrate the relic density allowed parameter space in the bi-dimensional plane of $\left[\Min-\mdm\right]$. The gray-shaded region corresponds to DM overabundance. The region corresponding to $k=0,\,q=1$ is shown via the lighter gray shade (visible in the left panel). As one can see, with the quantum effects into play, the region of overabundance is reduced. This can be explained by the fact that the PBH of a given mass now evaporates over a longer period of time, injecting entropy into the bath that dilutes the excess DM abundance. The upper and lower boundaries correspond to $\mdm>T_{\rm BH}^{\rm in}$ and $\mdm<T_{\rm BH}^{\rm in}$, respectively. The two boundaries show a discontinuity at PBH masses that correspond to $T_{\rm ev}=T_{\rm BBN}$ since we are considering all PBHs evaporate before the onset of BBN. A large part of the parameter space is in tension with the WDM bound [cf. Eq.~\eqref{eq:wdm}], as shown by the green-shaded region. This typically allows the heavier mass regime, depending on the choice of $k,\,q$ and $\Min$ in order to produce the right relic abundance. DM is produced only through gravity mediation, accounting for all the DM abundance that takes place along the black solid line.  Since to the left of such a line, DM is overproduced, that region coincides with the region of overabundance for DM Hawking radiated from PBHs.
\subsection{Production of baryon asymmetry}
\label{sec:bau}
We consider the vanilla leptogenesis scenario, where the out-of-equilibrium production of RHNs results in the generation of baryon asymmetry via lepton asymmetry. The relevant interaction Lagrangian is given by
\beq
 \mathcal{L} \supset -\frac12\, \sum_i M_{N_i}\, \overline{N_i^c}\, N_i - y_{N}^{ij}\, \overline{N_i}\, \widetilde{H}^\dagger\, L_j + {\rm h.c.}\,;~(i=1,\,2,\,3)\,,
\eeq
where the SM particle content is extended with the addition of three generations of right-handed neutrinos $N_i$, singlet under the SM gauge symmetry. The SM left-handed leptons doublets are identified as $L_i$ and $\widetilde{H}=i\,\sigma_{\rm 2}\,H^{\rm *}$ where $H$ represents the SM Higgs doublet. $\sigma_{\rm i}$ are the Pauli spin matrices. We assume the Majorana masses $M_{N_i}$ to be hierarchical $M_{N_1} \ll M_{N_{2,3}}$. We also consider lepton-number-violating interactions of $N_1$ are rapid enough to wash out the asymmetry originating from the decay of the other two. Therefore, only the CP-violating asymmetry from the decay of $N_1$ survives and is relevant for leptogenesis. Once right-handed neutrinos are produced from the evaporating PBHs, they can decay and result in a lepton asymmetry
\beq
Y_L=\frac{n_L}{s}=\epsilon_{\Delta L}\,\frac{n_{N_1}}{s}\,,
\eeq
where the CP asymmetry generated from $N_1$ decay reads~\cite{Davidson:2008bu}
\begin{align}\label{eq:cp-asym}
&\epsilon_{\Delta L} \equiv \frac{\Gamma_{N_1 \to \ell_i\, H } -\Gamma_{N_1 \to \bar\ell_i\, \bar H}}{\Gamma_{N_1 \to \ell_i\, H} + \Gamma_{N_1 \to \bar\ell_i\, \bar H}}
\simeq \frac{1}{8\, \pi}\, \frac{1}{(y_N^\dagger\, y_N)_{11}}\, \sum_{j=2, 3} \text{Im}\left(y_N^\dagger\, y_N\right)^2_{1j} \times \mathcal{F}\left(\frac{M_{N_j}^2}{M_{N_1}^2}\right),
\end{align}
with
\begin{equation}
    \mathcal{F}(x) \equiv \sqrt{x}\,\left[\frac{1}{1-x}+1-(1+x)\,\log\left(\frac{1+x}{x}\right)\right]\,.
\end{equation}
This can be further simplified to~\cite{Kaneta:2019yjn, Co:2022bgh}
\begin{equation}
    |\epsilon_{\Delta L}| \simeq \frac{3\, \delta_\text{eff}}{16\, \pi}\, \frac{M_{N_1}\, m_{\nu_i}}{v^2}\simeq 
    10^{-6} \delta_{\rm eff}\left(\frac{M_{N_1}}{10^{10}\,\rm GeV}\right)\left(\frac{m_{\nu_i}}{0.05~\rm{eV}}\right)\,,    
\end{equation}
where $i=2\,,3$ for normal hierarchy (a similar CP-asymmetry parameter can be obtained for the inverted hierarchy with $i=1\,,2$) and $\delta _{\rm eff}$ is the effective CP-violating phase
\begin{equation}
    \delta_\text{eff} = \frac{1}{(y_N)_{13}^2}\, \frac{\text{Im}(y_N^\dagger\,y_N)^2_{13}}{(y_N^\dagger\,y_N)_{11}}\,, 
\end{equation} 
whereas $v=174$ GeV is the Higgs vacuum expectation value. The produced lepton asymmetry is eventually converted to baryon asymmetry via electroweak sphaleron processes, leading to baryon number yield at the point of evaporation~\cite{Baumann:2007yr,Fujita:2014hha,Datta:2020bht}
\begin{align}
 Y_B (\aev)=&\,\frac{n_B}{s}\Big|_{\aev}=N_j\,\epsilon_{\Delta L}\,a_\text{sph}\,\frac{n_\text{BH} (\aev)}{s (\aev)}
\nonumber\\
=&\,Y_B^{\rm obs}\,\left(\frac{\delta_{\rm eff}}{1}\right)\,\left(\frac{m_\nu}{0.05\,{\rm eV}}\right)\,\left[2^k\,(2\,k+3)\right]^\frac{1}{2}\,q^{-\frac{2\,k+5}{2}}
\nonumber\\&\times
\begin{cases}
503.5\times\left(\frac{M_{N_1}}{10^{12}\,{\rm GeV}}\right)\,\left(\frac{M_P}{\Min}\right)^\frac{2\,k+1}{2} & \text{for}~M_{N_1}<T_{_{\rm BH}}^{\rm in}
\\[10pt]
2.97\times10^{15}\left(\frac{10^{12}\,{\rm GeV}}{M_{N_1}}\right)\,\left(\frac{M_P}{\Min}\right)^\frac{2k+5}{2} & \text{for}~M_{N_1}>T_{_{\rm BH}}^{\rm in}\,,
\end{cases}
\label{eq:yB}
\end{align}
where $a_{\rm sph}=28/79$ being the sphaleron conversion factor and $Y_B^{\rm obs}\simeq 8.7\times 10^{-11}$ is the observed baryon asymmetry~\cite{Planck:2018jri}.
\begin{figure}
    \centering
    \includegraphics[scale=.3]{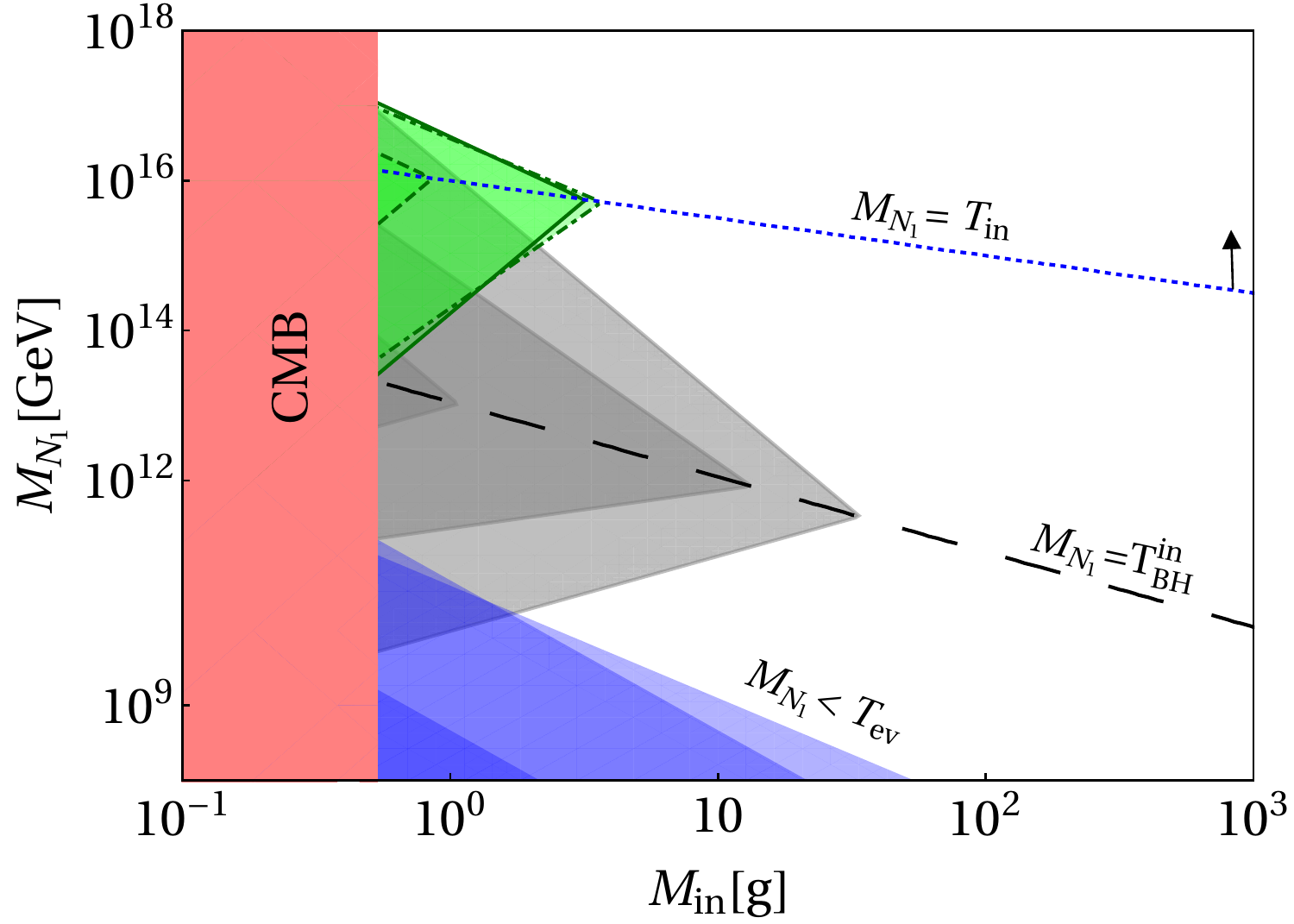}~\includegraphics[scale=.3]{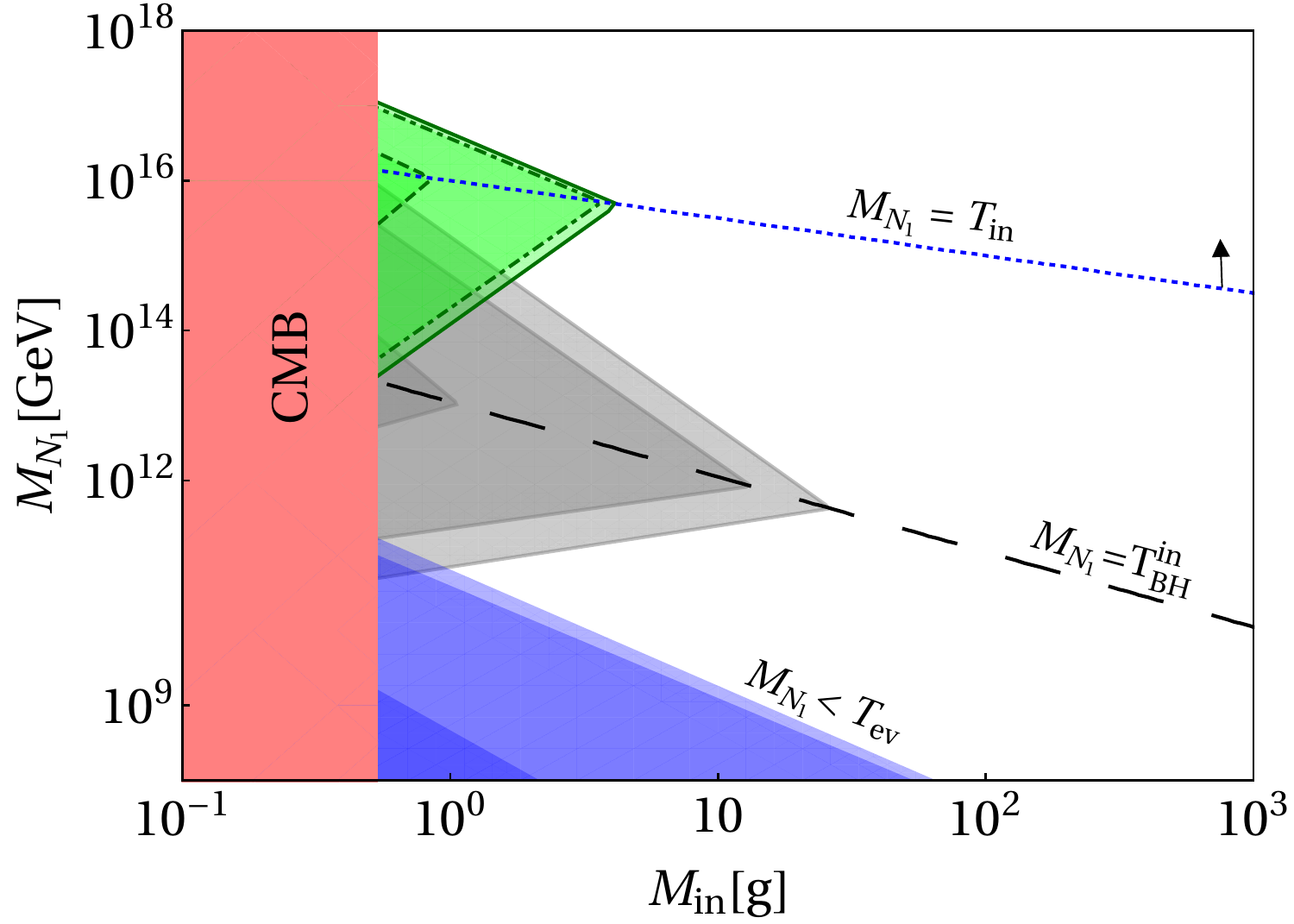}
    \caption{The gray shaded region from lighter to darker corresponds to the maximum allowed parameter space during PBH domination for $\{k,\,q\}=\{0.5,\,0.05\};\,\{0,\,1\};\,\{0.5,\,0.5\}$ (left panel) and $\{k,\,q\}=\{0.05,\,0.5\};\,\{0,\,1\};\,\{0.5,\,0.5\}$ (right panel). The green shaded region corresponds to the maximally allowed parameter space for leptogenesis, where we only consider gravity-mediated production of the RHNs from the bath for $\{k,\,q\}=\{0.5,\,0.05\};\,\{0,\,1\};\,\{0.5,\,0.5\}$ in the left panel, and $\{k,\,q\}=\{0.05,\,0.5\};\,\{0,\,1\};\,\{0.5,\,0.5\}$ in the right panel, shown via dot-dashed, solid and dashed boundaries respectively.}
    \label{fig:ybplt}
\end{figure}
Following subsection~\ref{sec:grav}, we see that right baryon asymmetry can also be produced from the CP-violating decay of the RHNs that are produced from the thermal bath mediated by gravity. In this case, the final asymmetry reads
\begin{align}
& Y_B^{\rm obs}=Y_0^N\,\epsilon_{\Delta L}\,a_{\rm sph}\times \frac{S(T_\text{in})}{S(T_\text{ev})}\,,     
\end{align}
where $Y_0^N$ is given by Eqs.~\eqref{eq:grav-yield1} and \eqref{eq:grav-yield2}, with $m_j\to M_{N_1}$.

The gray shaded region in Fig.~\ref{fig:ybplt} shows the maximally available parameter space satisfying the observed baryon asymmetry, corresponding to $\delta_{\rm eff}=1$. In either panel, the intermediate gray region corresponds to the $k=0,\,q=1$ case. Note that, for the same choice of $k$, an increase in $q$ reduces the resulting parameter space. This is because, for the same PBH mass, a larger $q$ results in a longer lifetime, hence a longer period of entropy injection and corresponding dilution of the asymmetry. An important constraint on the parameter space arises from the fact that if $M_{N_1}<T_{\rm ev}$, then the RHNs produced from PBH evaporation are in a therm bath, and then washout processes are in effect. It is worth mentioning here for $k=0.03$ and $q=0.05$, it is possible to obtain a viable parameter space for leptogenesis with $\Min\leq 10^{3}$ g. Hence, to ensure non-thermal production of baryon asymmetry, one must follow $M_{N_1}>T_{\rm ev}$. This forbids the blue-shaded region. 
\begin{figure}
    \centering
   \includegraphics[scale=0.5]{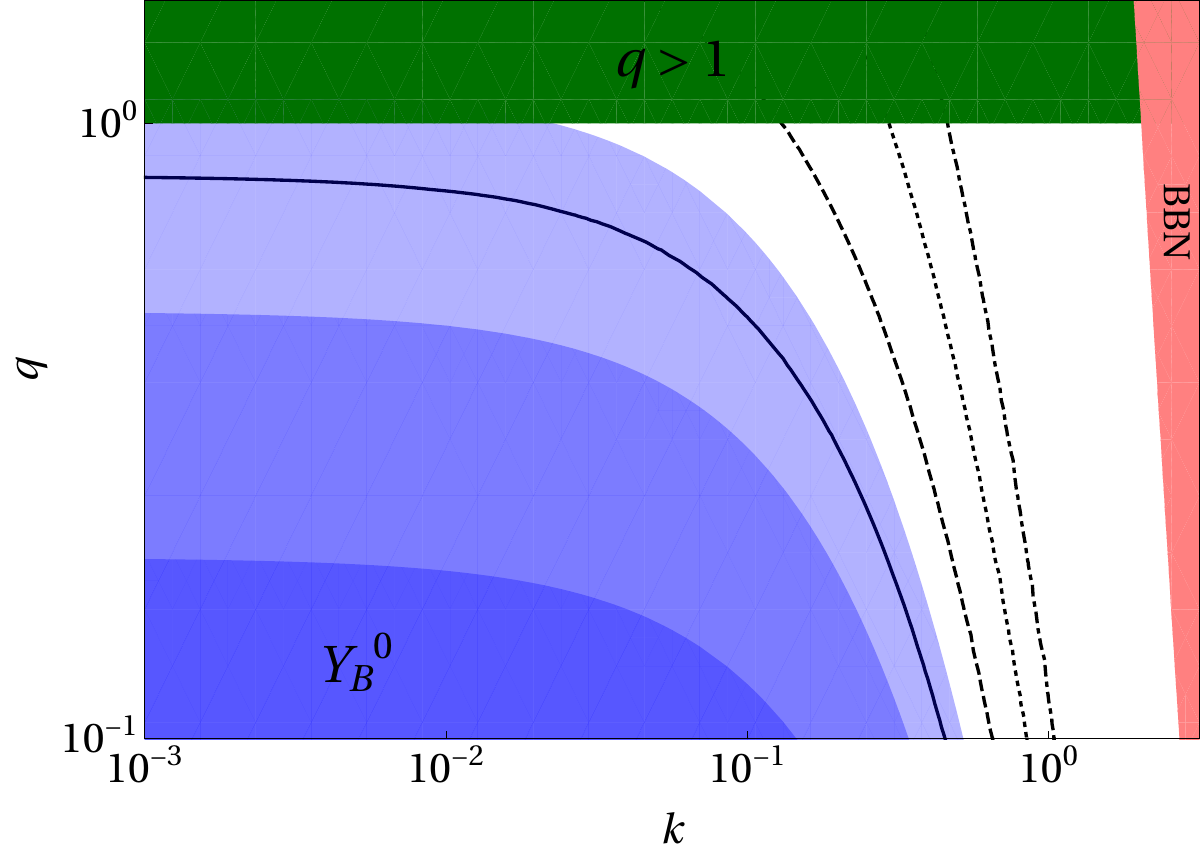}
    \caption{The blue shaded region corresponds to observed BAU for $M_{N_1}=\{10^{12},\,10^{13},\,10^{14}\}$ GeV shown via lighter to darker shades of blue. Different contours satisfy total DM abundance for a scalar DM with mass $\mdm=\{0.1,\,1,\,10,\,100\}$ MeV (shown via solid, dashed, dotted, and dot-dashed pattern respectively) produced from PBH evaporation. We consider $\Min=10$ g. The red-shaded region is disallowed from BBN bound on PBH lifetime, while the green-shaded region is forbidden as $q>1$. The whole DM parameter space is discarded from the WDM limit, which is not shown for clarity of the figure.}
    \label{fig:cogen}
\end{figure}
As before, observed baryon asymmetry can also be produced due to the CP-violating decay of the RHNs emerging from the thermal bath, mediated by massless graviton. In this case, the maximum parameter space allowed is shown in the green-shaded region. Note that, in order to avoid washout due to inverse decay, one needs to have $M_{N_1}>\Trh\equiv T_{\rm in}$, as indicated by the arrowhead along the blue-dashed line corresponding to $M_{N_1}=T_{\rm in}$. A larger $q$, hence a larger entropy dilution, also results in reduced parameter space for asymmetry produced via graviton exchange, as evident from the green shaded region with different boundary patterns.

We depict a common parameter space in $[q-k]$ plane that yields the correct abundance for scalar dark matter (indicated by the black contours) and also gives rise to the observed baryon asymmetry (represented by the blue shaded region) in Fig.~\ref{fig:cogen}. While we observe that it is conceivable to achieve a simultaneous solution for both DM abundance and the baryon asymmetry with a DM mass of 0.1 MeV, all parameter regions allowed for DM relic density are invalidated by the warm DM constraint. We have selected the PBH mass to be $\Min=10$ g to ensure the generation of the correct baryon asymmetry [cf. Fig.~\ref{fig:ybplt}]. For DM of mass $\gtrsim 10^{12}$ GeV, although the WDM limit can be somewhat relaxed [cf. Fig.~\ref{fig:relic}], but this necessitates massive PBHs $\gtrsim 10^5$ g, which renders the correct baryon asymmetry unattainable. Consequently, we conclude that for a given PBH mass, a parameter space where both the correct DM abundance and BAU can coexist is not achievable within the present framework.  
\section{Dark radiation from memory-burdened PBH}
\label{sec:DR}
As we know, once PBHs are evaporated, along with the SM radiation, beyond the SM states, like the DM or the RHNs, can also be produced via Hawking radiation. Should there be additional light states with feeble coupling to the SM, known as ``dark radiation" (DR), these particles will emerge through Hawking radiation, thereby influencing the radiation density of the universe during BBN and recombination. The contribution of the DR component to the effective number of relativistic degrees of freedom (DoF) is parametrized via
\begin{align}
& \Neff = \frac{\rho_{\rm DR}(T_{\rm eq})}{\rho_R(T_{\rm eq})}\,\left[N_\nu+\frac{8}{7}\,\left(\frac{11}{4}\right)^{4/3}\right]\,,    
\end{align}
where $T_{\rm eq}\simeq 0.75$ eV is the temperature at the late radiation-matter equality. Within the SM, taking the non-instantaneous neutrino decoupling into account, one finds $N_\nu = 3.044$~\cite{Dodelson:1992km, Hannestad:1995rs, Dolgov:1997mb, Mangano:2005cc, deSalas:2016ztq, EscuderoAbenza:2020cmq, Akita:2020szl, Froustey:2020mcq, Bennett:2020zkv}. Here, $\rho_{\rm DR}$ is the energy density of the dark radiation.  Following Refs.~\cite{Hooper:2019gtx} and \cite{Masina:2020xhk}, we see,
\begin{align}
& \DNeff = \frac{\rho_{\rm DR}(\Tev)}{\rho_R(\Tev)}\,\frac{\gss(T_{\rm eq})}{g_\star(T_{\rm eq})}\,\left[\frac{\gss(T_{\rm eq})}{\gss(\Tev)}\right]^\frac{1}{3}\,\left[N_\nu+\frac{8}{7}\,\left(\frac{11}{4}\right)^{4/3}\right]\,.    
\end{align}
In the case of PBH domination, this expression simplifies to
\begin{align}
& \DNeff =  \frac{g_{{\rm DR},H}}{g_{\star,H}}\,\frac{\gss(T_{\rm eq})}{g_\star(T_{\rm eq})}\,\left[\frac{\gss(T_{\rm eq})}{\gss(\Tev)}\right]^\frac{1}{3}\,\left[N_\nu+\frac{8}{7}\,\left(\frac{11}{4}\right)^{4/3}\right]\simeq 0.07\times\frac{g_{{\rm DR},H}}{4}\,\frac{106}{g_{\star,H}}\,,    
\end{align}
where we have considered $k=0.1,\,q=0.5$ and $\Min=10$ g. Here, $g_{{\rm DR},H}=\{4,\,2\}$ for a Dirac and a Weyl fermion respectively, $g_{{\rm DR},H}=1.82$ for a scalar boson, $g_{{\rm DR},H}=1.23$ for a massive vector boson and $g_{{\rm DR},H}=0.05$ for a massless graviton [cf. Eq.~\eqref{eq:gstTBH}]. Present CMB measurement from Planck legacy data~\cite{Planck:2018jri} provides $\DNeff\simeq 0.34$. On inclusion of baryon acoustic oscillation (BAO) data, the constraint becomes more stringent: $N_\text{eff} = 2.99 \pm 0.17$. Upcoming CMB experiments like CMB-S4~\cite{Abazajian:2019eic} and CMB-HD~\cite{CMB-HD:2022bsz} will be sensitive to a precision of $\DNeff \simeq 0.06$ and $\DNeff \simeq 0.027$, respectively.
\begin{figure}
    \centering   \includegraphics[scale=.36]{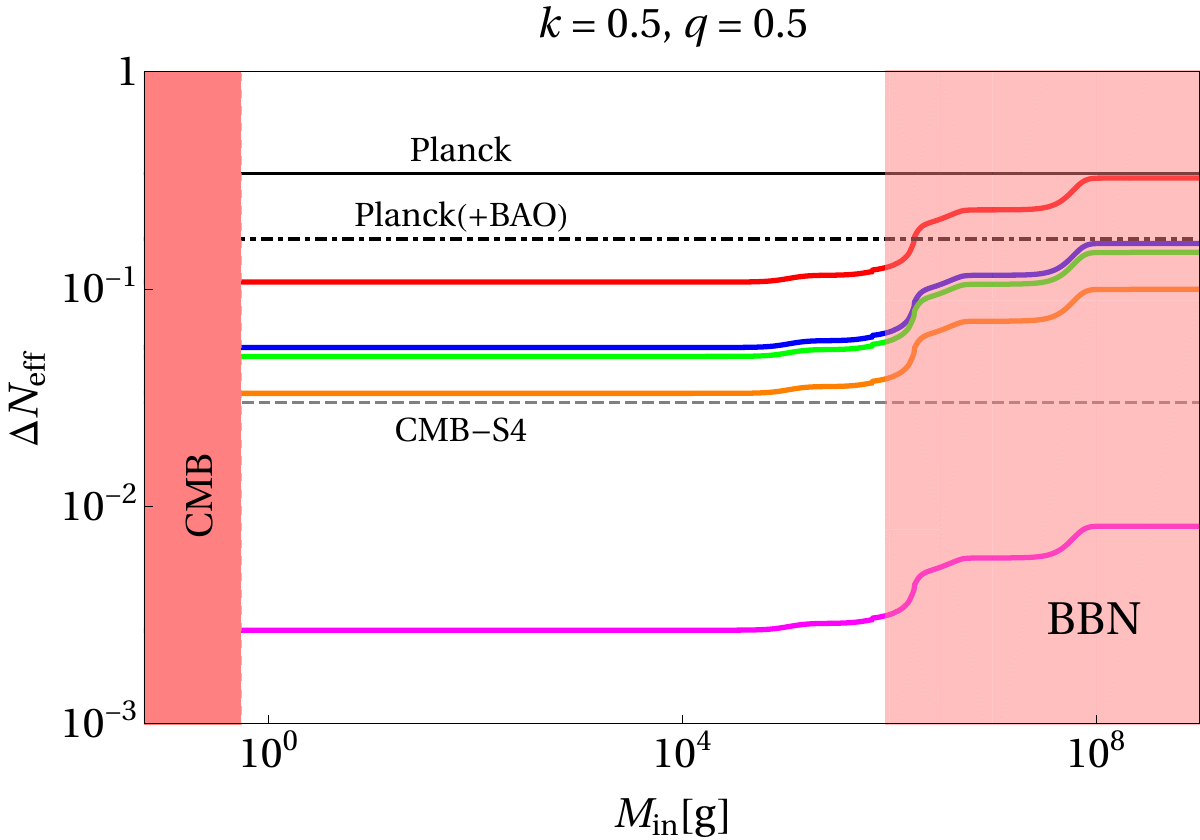}~~\includegraphics[scale=.36]{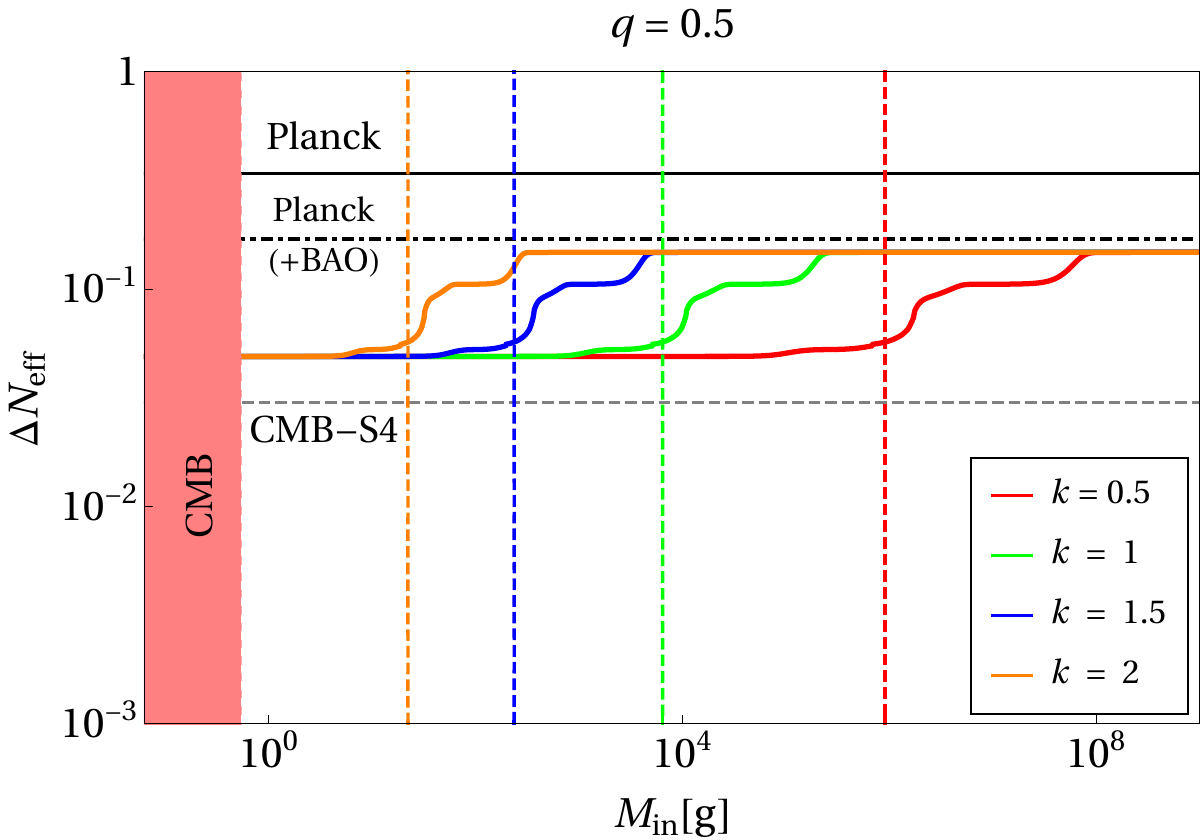}
    \caption{Contribution of dark radiation to $\DNeff$, produced from PBH evaporation. In the left panel, we show contribution due to a spin-1/2 Dirac fermion (red), spin-1/2 Weyl fermion (blue), spin-0 boson (green), spin-1 massive vector boson (orange), and massless spin-2 graviton (magenta), for  $k=q=0.5$. The red-shaded region corresponds to the PBH mass that evaporates after BBN. In the right panel, we illustrate the contribution to $\DNeff$ due to a spin-0 scalar boson for different choices of $k$ at a fixed $q=0.5$. For each $k$, the vertical dashed line shows the contour for $t_{\rm ev}=t_{\rm BBN}\simeq 1$ sec. Also shown are the present constraints from Planck and future CMB-S4 bound on $\DNeff$ via black solid and gray broken horizontal lines, respectively.}
    \label{fig:DRplt}
\end{figure}

In the left panel of Fig.~\ref{fig:DRplt}, we show contributions from different spins to DR, produced from PBH evaporation to $\DNeff$. Here, we choose $k=q=0.5$ as a benchmark value. Due to the highest number of degrees of freedom, a Dirac fermion (in red) has a maximum contribution to $\DNeff$, while a graviton (in magenta) contribution is minimum. As we see, PLANCK+BAO data already excludes Dirac fermion as a viable DR candidate for PBH masses $\Min\gtrsim 10^7$ g. Future experiments like CMB-S4 can effectively rule out all spins, leaving only graviton as a possible candidate for DR. In the right panel, we illustrate the effect of memory burden on $\DNeff$ for a fixed $q$, with different choices of $k$, considering a spin-0 boson contributing to the DR.
In general, we expect Hawking evaporation to generate a sizable contribution to $\DNeff$ only if the early universe included a PBH-dominated era or if there exists a large number of light decoupled
species, i.e., $g_{{\rm DR},H}\gg 1$ (the typical example being the supersymmetric (SUSY) models). 
\section{Induced gravitational wave from PBH density fluctuations}
\label{sec:gw}
PBHs can play as a source for primordial GWs in several ways, e.g., by inducing large curvature perturbation that is responsible for its formation~\cite{Baumann:2007zm,Espinosa:2018eve,Domenech:2019quo,Ragavendra:2020sop,Inomata:2023zup,Franciolini:2023pbf,Firouzjahi:2023lzg,Heydari:2023xts,Heydari:2023rmq,Maity:2024odg}, by radiating gravitons that constitute high-frequency GW~\cite{Fujita:2014hha}, through PBH-merger~\cite{Zagorac:2019ekv} or by the fluctuation of PBH number density~\cite{Domenech:2020ssp,Papanikolaou:2020qtd,Domenech:2021wkk,Papanikolaou:2022chm,Bhaumik:2022pil,Bhaumik:2022zdd,Balaji:2024hpu}. Here we consider the last possibility. Our final goal is to establish a connection between the scale of (PBH-induced) leptogenesis with the GW amplitude and frequency, such that any potential detection of GW can be interpreted as a probe for the scale of leptogenesis itself. 

We consider a gas of PBHs having all the same mass randomly distributed in space, with a Poissonian spatial distribution~\cite{Papanikolaou:2020qtd}. These inhomogeneities in the distribution of PBHs lead to density fluctuations, which are isocurvature in nature. When PBHs become the dominant component of the Universe's energy density, these isocurvature perturbations transition into adiabatic perturbations, which, in the second order, can generate GWs. Here, we closely follow Refs.~\cite{Papanikolaou:2020qtd,Domenech:2020ssp,Balaji:2024hpu} in estimating the modification to the curvature fluctuation due to the effect of the memory burden on the evaporation process. Considering a generic power spectrum in wavenumber $K$ of fluctuations in gravitational potential $\Phi$ during the PBH-dominated stage can be expressed as
\bea
\mathcal{P}_\Phi (K)=\mathcal{A}_\Phi \left(\frac{K}{K_{\rm UV}}\right)^n\Theta(K_{\rm UV}-K)\,,
\eea
where $\mathcal{A}_\Phi$ is the amplitude of the power spectrum with index $n$, and $K_{\rm UV}$ represents the wavenumber corresponding to the cut-off scale. One can then calculate the induced GW spectrum for a given primordial spectrum of fluctuations by multiplying with the appropriate kernel and integrating across the internal momenta, as can be found in Refs.~\cite{Papanikolaou:2020qtd,Domenech:2020ssp,Balaji:2024hpu}. If we assume that the PBH decay process is instantaneous, the peak amplitude of the induced GWs at the evaporation end can be written as \cite{Balaji:2024hpu}

\bea\label{eq:omegapeak}
\Omega_{\rm GW,\,ev}^{\rm peak}\simeq \frac{q^4}{24576\sqrt{3}\pi}\left(\frac{3+2k}{3\,\sqrt{2(1+k)}}\right)^{-\frac{4}{3+2k}}\,\left(\frac{K_{\rm eq}^{\rm BH}}{K_{\rm UV}}\right)^8\left(\frac{K_{\rm UV}}{K_{\rm ev}}\right)^{7-\frac{4}{3+2k}}\,,
\eea
where $K_{\rm eq}^{\rm BH}$ is the wave number associated with the early radiation-matter equality. Now, let us calculate all the relevant wave numbers that we need to estimate the final GW spectrum.\\
The description of the PBH gas in terms of a continuous fluid is only valid at scales larger than the mean separation distance $D_{\rm mean}$, which imposes an ultra-violet (UV) cutoff in the power spectrum~\cite{Papanikolaou:2020qtd}
\begin{align}\label{eq:kUV}
K_{\rm UV} = \frac{\ain}{D_{\rm mean}} = \ain\,\left(\frac{4\,\pi\,\rho_\text{BH}\left(T_\text{in}\right)}{3\,\Min}\right)^{1/3} = K_{\rm in}\,\gamma^{-1/3}\,\beta^{1/3}\,,
\end{align} 
where $K_{\rm in}=\ain\,\hin$ is the wave number defined at the formation point. Utilizing the fact that the comoving PBH number density is conserved, the momentum mode $K_{\rm ev}=\aev \Hev$ re-entering at the evaporation time can be expressed as
\begin{align}\label{Eq:kev}
    K_{\rm ev}&=\left(\frac{\Hev}{\hin}\right)^{1/3}K_{\rm in}\,\beta^{1/3}\,q^{1/3}
    =\left(\frac{2^k\,(3+2k)\,\epsilon}{6\pi\gamma}\right)^{1/3}\left(\frac{\Mpl}{q\,\Min}\right)^{\tfrac{2}{3}(1+k)}K_{\rm in}\,\beta^{1/3}.
\end{align}
Combining these expressions, we obtain \footnote{Note that our expression for $K_{\rm UV}/K_{\rm ev}$ differs from Ref.\cite{Balaji:2024hpu} with a factor $q^{1/3}$, which they claim will be present when memory effect takes place during PBH domination, i.e., $t_q>t_{\rm eq}^{\rm BH}$. However, we found that $K_{\rm UV}/K_{\rm ev}$  remains the same for both $t_q>t_{\rm eq}^{\rm BH}$ and $t_q<t_{\rm eq}^{\rm BH}$.}
\begin{align}\label{eq:kfrac}
    \frac{K_{\rm UV}}{K_{\rm ev}}\approx 4133\,e^{8k}\,\left(\frac{3}{3+2k}\right)^{\frac{1}{3}}\left(\frac{q\,\Min}{1\,{\rm g}}\right)^{\tfrac{2}{3}(1+k)}\,.
\end{align}
The frequency $f_{\rm ev}=a_{\rm ev}\Hev/(2\pi)$ can be estimated as 
\begin{align}\label{eq:fev}
    f_{\rm ev}&=\frac{\sqrt{g_\star(\Tev)}}{6\sqrt{10}\Mpl}\left(\frac{\gss(T_0)}{\gss(\Tev)}\right)^{1/3}\Tev\,T_0\simeq 963\,{\rm Hz}\,e^{-12k}\,\sqrt{\frac{3+2k}{3}}\,\left(\frac{q\,\Min}{1\,{\rm g}}\right)^{-\left(\tfrac{2}{3}+k\right)},
\end{align}
which leads to 
\begin{align}\label{eq:fuv}
    f_{\rm UV}\simeq 4.8\times 10^{6}\,{\rm Hz}\,e^{-4k}\,\left(\frac{3+2k}{3}\right)^{\frac{1}{6}}\,\left(\frac{q\,\Min}{1\,{\rm g}}\right)^{-\left(\tfrac{5}{6}+\tfrac{k}{3}\right)}\,,
\end{align}
implying that the modified Hawking radiation typically leaves a high-frequency GW signature in the case of PBH reheating. The expression for the wave number $k_{\rm eq}^{\rm BH}$ associated with the early radiation-matter equality solely depends on the fact that in which domination memory burden effect starts, which is decided by a $\beta$ value, $\beta_\star$. Therefore, for $\beta>\beta_\star$, the memory burden effect starts during PBH domination. 
\begin{figure}
    \centering   
    \includegraphics[scale=0.36]{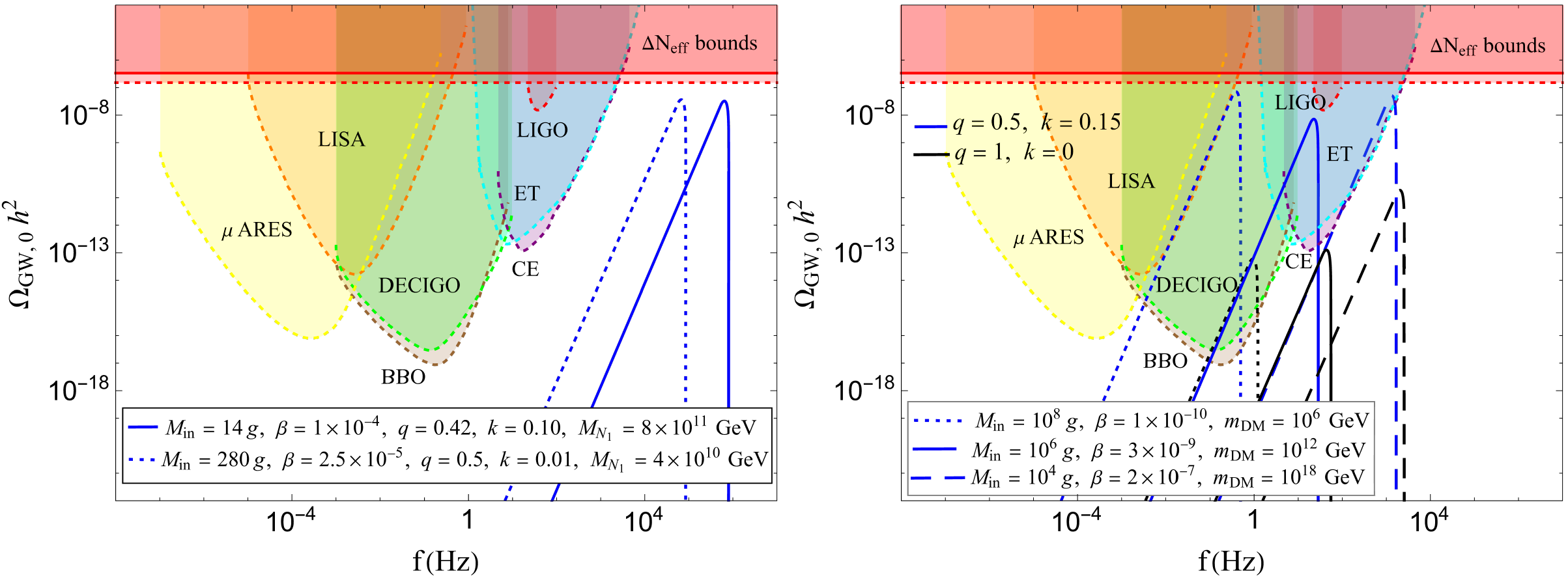}
    \caption{Induced GW spectrum at the present epoch, from PBH density fluctuations. We project future sensitivity curves from several low and high-frequency GW detectors, along with BBN bound on $\DNeff$ from present and future experiments, shown via the horizontal lines. In the right panel, we also compare the spectrum of induced GWs for $k=0,\,q=1$ scenario, as shown via the black lines.}
    \label{fig:GWs}
\end{figure} 
 The time corresponding to the early radiation-PBH equality, $t_{\rm eq}^{\rm BH}$, obtained through the condition $\rho_{\rm BH}(t_{\rm eq}^{\rm BH})=\rho_R(t_{\rm eq}^{\rm BH})$, leads to
\begin{align}
\beta_\star=\sqrt{\frac{3\,\epsilon}{16\,\pi\,\gamma\,(1-q^3)\,S(\Min)}}\simeq 8.2\times 10^{-6}\,\left(\frac{1\,\text{g}}{\Min}\right)\,\left[1-\left(\frac{q}{0.5}\right)^3\right]^{-1/2}\,,
\end{align}
under the condition $t_{\rm eq}^{\rm BH}=t_q$, where $S(\Min)$ is given by Eq.~\eqref{eq:TbhS}. 
Now the mode $K_{\rm eq}^{\rm BH}=a_{\rm eq}^{\rm BH} H_{\rm eq}^{\rm BH}$ can be written in terms of $K_{\rm in}$ as
\begin{align}\label{eq:keqBH}
    K_{\rm eq}^{\rm BH}&=\begin{cases}
        \sqrt{2}\,\beta\,K_{\rm in}, \quad{\rm for}\quad \beta>\beta_\star,\\[10pt]   \sqrt{2}\,q\,\beta\,K_{\rm in}, \quad{\rm for}\quad \beta<\beta_\star\,, 
    \end{cases}
\end{align}
Note that $\sqrt{2}$ factor arises from the fact that at equality, the Hubble parameter has equal contribution from matter (PBH) and radiation energy densities. The induced GWs spectrum as a function of $f$ can be written in terms of peak value as~\cite{Balaji:2024hpu}
\begin{align} \label{eq:GW}
    \Omega_{\rm GW,ev}(f)&\approx\Omega_{\rm GW,ev}^{\rm peak}\left(\frac{f}{f_{\rm UV}}\right)^{\frac{11+10k}{3+2k}}\,\mathcal{I}(f,k)\,,
\end{align}
where 
\begin{align}
    \mathcal{I}(f,k)&=\int_{\xi_0(f)}^{\xi_0(f)}ds\,(1-s^2)^2\,(1-c_s^2s^2)^{-(1+\frac{2}{3+2k})}\,,
\end{align}    
with
\begin{align}
& \xi_0(f)=\begin{cases}
        1,\quad{\rm for}\quad \frac{f_{\rm UV}}{f}\,\geq\, \frac{1+c_s}{2c_s},\\[10pt]
        \frac{2\,f_{\rm UV}}{f}-\frac{1}{c_s},\quad{\rm for}\quad \frac{1+c_s}{2c_s}\,\geq\, \frac{f_{\rm UV}}{f}\, \geq\, \frac{1}{2c_s},\\[10pt]
        0,\quad{\rm for}\quad \frac{1}{2c_s}\,\geq\, \frac{f_{\rm UV}}{f}\,,
    \end{cases}    
\end{align}
where $c_s=1/\sqrt{3}$ is the sound speed during the radiation-dominated epoch. To find the above expression, we use the index of the power spectrum $n=-\frac{5}{3}+\frac{4\,k}{9+6k}$. Upon substitution of the expression related to the wave numbers, Eqs. (\ref{eq:kfrac}), (\ref{eq:fuv}), and (\ref{eq:keqBH}) into (\ref{eq:omegapeak}), we have 
\begin{align}\label{eq:omega-peak-ev}
& \Omega_{\rm GW,ev}^{\rm peak}\simeq \frac{4133^{-\frac{4}{3+2k}}\,q^4}{2.3\times10^{-20}}\left(\frac{3+2k}{3}\right)^{-\frac{7}{3}+\frac{4}{9+6k}}\,\beta^{\frac{16}{3}}\,e^{8\,k\,(7-\frac{4}{3+2k})}
\nonumber\\&\times 
\left(\frac{q\,\Min}{1\,{\rm g}}\right)^{\frac{2}{3}(1+k)\left(7-\frac{4}{3+2k}\right)}
\begin{cases}
1 & \text{for}\quad \beta>\beta_\star
\\[10pt]
q^8 & \text{for}\quad \beta<\beta_\star\,.
\end{cases}
\end{align}
The peak of the amplitude is exponentially enhanced with $k$, but also sensitive to $\Min$ and $\beta$. We thus obtain the present-day GW spectral energy density as 
\bea
\Omega_{{\rm GW},0}\,h^2(f)&\simeq c_g ~\Omega_{\text{rad},0}\,h^2~\,\Omega_{{\rm GW},{\rm ev}}(f)=1.62\times 10^{-5}\,\Omega_{{\rm GW},{\rm ev}}(f)\,h^2 ,
\eea
where the present-day radiation energy density is $\Omega_{\text{rad},0}\,h^2=4.16\times 10^{5}$ (i.e., including photons and all three species of neutrinos) and $c_g =\l(g_k/g_0\r) \left(g_{\mathrm{s}0}/g_{\mathrm{s}k}\right)^{4/3}\simeq 0.4$,
with $(g_k,g_{\mathrm{s}k})$ and $(g_0,g_{\mathrm{s}0})$ are the 
effective number of relativistic degrees of freedom associated with the radiation energy density and entropy when the wave number $K$ re-enters the horizon and today, respectively. Note that for $\beta>\beta_\star$, where the memory burden effect starts during PBH domination, our analysis is only applicable for $q>0.41$~\cite{Balaji:2024hpu}. Otherwise, another intermediate radiation domination will arise during PBH domination, where our analysis fails to capture those possibilities. 

In Fig.~\ref{fig:GWs}, are displayed the projected sensitivity reach of several proposed GW experiments such as $\mu$-ARES~\cite{Sesana:2019vho}, LISA~\cite{LISA:2017pwj}, Einstein Telescope (ET)~\cite{Punturo:2010zz, Hild:2010id, Sathyaprakash:2012jk, Maggiore:2019uih}, Cosmic Explorer~\cite{LIGOScientific:2016wof, Reitze:2019iox} the Big Bang Observer (BBO)~\cite{Crowder:2005nr, Corbin:2005ny, Harry:2006fi}, and  DECIGO ~\cite{Kawamura_2011, Kawamura:2019jqt}. Using the present CMB measurement from Planck legacy data~\cite{Planck:2018jri}, we find $\Omega_{{\rm GW},0}\,h^2\lesssim 2\times 10^{-6}$, with $\DNeff\simeq 0.34$. We show the present bound from Planck and future projection from CMB-S4 by the red solid and red dashed horizontal lines, respectively. We notice the induced GW spectrum has frequency dependency $\propto f^{\frac{11+10\,k}{3+2k}}$ [cf. Eq.~\eqref{eq:GW}] and the peak frequency is associated with the frequency $f_{\rm UV}$. Moreover, the spectrum drops sharply above the frequency scale $2c_s\,f_{\rm UV}$ and goes to zero at the cutoff scale $2f_{\rm UV}$. Since $f_{\rm UV}\propto K_{\rm in}$ and smaller mass BHs are formed earlier, thus with the decreases in the PBH mass, the peak frequency shifts more towards the end of the inflation, or in other words, to higher frequencies. In the right panel, we compare the corresponding GW spectrum from PBH density fluctuation for $k=0,\,q=1$, i.e., without taking the effect of memory burden into account. Here, we see that not only is the peak of the spectrum shifted to higher frequencies, but the peak amplitude is substantially suppressed. This is because for $k=0$, there is no enhancement in the peak amplitude, as evident from Eq.~\eqref{eq:omega-peak-ev}. Also, in the absence of memory effect, $f_{\rm UV}$ has no exponential suppression [cf. Eq.~\eqref{eq:fuv}], as a consequence, in this case, the cut-off frequency shifts towards a larger magnitude compared to the scenario with $k>0$. We can, therefore, infer that inclusion of backreaction effects can actually enhance the detection potential for the induced GWs due to PBH density fluctuation.

Both the GW energy density and the DM/BAU parameter space are controlled by the common set of parameters: $\{k,\,q\}$ once we fixed the PBH parameters such as $M_{\rm in}$ and $\beta$. In the case of PBH domination for a given $k,\,q$, and $M_{\rm in}$ as we have seen before, one can choose the DM and RHN mass such that correct relic abundance and baryon asymmetry are produced. Here, we show that such combinations of $k,\,q$, and PBH parameters can potentially give rise to a detectable GW signal. A noteworthy feature here is the fact that parameters corresponding to the right BAU produce GW in the kHz-MHz frequency, while those that provide the right amount of DM give rise to GW ranging from mHz-kHz frequency. This is because, to produce right baryon asymmetry from PBH evaporation, typically very lighter PBHs $M_{\rm in} <10^3$ g is required, for which $f_{\rm UV}$ is higher, following Eq.~\eqref{eq:fuv}. On the other hand, the production of observed DM abundance requires much heavier PBHs, $M_{\rm in}>10^3$ g, which leads to lower $f_{\rm UV}$. Let us recall that the value of $k$ may be extracted by precisely measuring the slope of the spectrum, whereas $\Min$ may be extracted from the peak frequency. In summary, future detection of GW in either of these two frequency regimes may indicate the possibility of the production of cosmic relics, namely DM and baryon asymmetry, from the evanescence of a memory-affected PBH.      
\section{Conclusions}
\label{sec:concl}
In this study, we have investigated the potential solutions to two enduring challenges in particle-cosmology: elucidating the abundance of dark matter (DM) and the dynamic generation of baryon asymmetry (BAU), solely through the evaporation process of primordial black holes (PBHs). Specifically, we have accounted for the impact of backreaction resulting from the emitted quanta during PBH evaporation, which gives rise to the so-called ``memory burden" effect. This effectively diminishes the rate of PBH evaporation by inverse powers of the black hole entropy. Typically, when the energy of the emitted quanta reaches a level comparable to that of the PBH itself, Hawking's semi-classical approach becomes inadequate, and the memory effect becomes significant. Given that for PBHs with masses greater than the Planck mass ($M_P$), the black hole entropy is considerable; the memory burden can substantially prolong the lifespan of the PBH [cf. Fig.~\ref{fig:gammak} and Fig.~\ref{fig:Min-k}].

The inclusion of backreaction increases the viable parameter space for DM produced form PBH evaporation, that can make up all of the abundance [cf. Fig.~\ref{fig:relic}]. However, the observed DM relic can only be produced from PBHs with masses $\gtrsim 10^3$ g, satisfying bound on warm DM due to Ly-$\alpha$. On the other hand, generation of baryon asymmetry via vanilla leptogenesis requires light PBHs $\lesssim 10^3$ g with corresponding right-handed neutrinos (RHN) having masses $\gtrsim 10^{11}$ GeV [cf. Fig.~\ref{fig:ybplt}]. Consequently, obeying all existing constraints from BBN, CMB, and WDM while simultaneously satisfying DM relic and BAU is not possible in the simplest framework [cf. Fig.~\ref{fig:cogen}]. In addition to the production from PBH evaporation, we also consider another inherent source of particle generation, namely, gravitational UV freeze-in. In this case, the cosmological relics are produced from the 2-to-2 scattering of the bath particles, mediated by graviton. Together, we explore the viable parameter space for DM and baryon asymmetry with a purely gravitational origin.

Finally, we discuss the detectability of primordial gravitational waves (GW), originating from the induced PBH number density fluctuations. We find, future GW detectors can indeed probe such a spectrum   [cf. Fig.~\ref{fig:GWs}], typically in the frequency range of mHz-MHz. Importantly, any detection of such spectrum paves the way to detect modified Hawking radiation, thereby constraining either the DM mass scale or the scale of leptogenesis.
\section*{Acknowledgment}
MRH wishes to acknowledge support from the Science and Engineering Research Board (SERB), Government of India (GoI), for the SERB National Post-Doctoral fellowship, File Number: PDF/2022/002988. O.Z. has been partially supported by Sostenibilidad-UdeA, the UdeA/CODI Grants 2022-52380 and  2023-59130, and Ministerio de Ciencias Grant CD 82315 CT ICETEX 2021-1080. The authors thank Guillem Domènech for useful communications, and Debasish Borah for providing comments on the manuscript.

\bibliographystyle{JHEP}
\bibliography{Bibliography}
\end{document}